\documentclass[onecolumn,showpacs,preprintnumbers,amsmath,amssymb,nofootinbib,12pt]{revtex4}
\usepackage{graphicx}
\usepackage{dcolumn}
\usepackage{bm}

\input psfig.sty

\newcommand{\bq}{\begin{equation}}
\newcommand{\eq}{\end{equation}}
\newcommand{\bqn}{\begin{eqnarray}}
\newcommand{\eqn}{\end{eqnarray}}

\newcommand{\lb}{\label}

\def\gappr{\lower 3pt\hbox{$\buildrel > \over \sim\;$}}
\def\gappl{\lower 3pt\hbox{$\buildrel < \over \sim\;$}}
\def\limiter{\lower 7pt\hbox{$\buildrel{\textstyle\longrightarrow}\over{\scriptscriptstyle ~~s\rightarrow\infty~~}\;$}}
\def\ablim{\lower 9pt\hbox{$\buildrel{\textstyle\longrightarrow}\over{\scriptscriptstyle ~~~a\rightarrow b~~~}\;$}}
\def\x0lim{\lower 11pt\hbox{$\buildrel{\textstyle\longrightarrow}\over{\scriptscriptstyle ~~x^0\rightarrow-\infty~~}\;$}}
\def\xlim{\lower 8.5pt\hbox{$\buildrel{\textstyle\longrightarrow}\over{\scriptscriptstyle ~~x_\pm\rightarrow-\infty~~}\;$}}
\def\pdot{\raise 1.5pt\hbox{.}}

\def\dal{\hbox{$\sqcup$\hbox to 0pt{\hss$\sqcap$}}}

         \def\ex{{\rm e}}
         \def\dal{\hbox{$\sqcup$\hbox to 0pt{\hss$\sqcap$}}}

\begin{document}
            
\title{\Large Strange stars properties calculated in the\\
framework of the Field Correlator Method}
\author{F. I. M. Pereira}\email{flavio@on.br}
\affiliation{Observat\'orio Nacional, MCT, Rua Gal. Jos\'e Cristino 77, 
20921-400 Rio de Janeiro RJ, Brasil}

\date{\today}

\begin{abstract}
  We calculate the strange star properties in the framework of 
the Field Correlator Method. 
  We find that for the values of the gluon condensate $G_2=0.006\;{\rm GeV}^4$ and 
$G_2=0.0068\;{\rm GeV}^4$, which give a critical temperature $T_c\sim170\;{\rm MeV}$
at $\mu_c=0$, the sequences of strange stars are compatible with some of the 
semi-empirical mass-radius relations and data obtained from astrophysical observations.
\end{abstract}


\maketitle


\section{\bf Introduction}
\lb{int}

 In the last decades, a great effort has been made to understand the properties of 
nuclear matter at densities higher than nuclear densities.
 From heavy ion collisions and astrophysical observations of compact objects, many 
attempts have been made to determine the equation of state (EOS) for dense nuclear 
matter in both hadronic and quark phases. 
  In relativistic heavy ions experiments (RHIC and LHC in the near future), extreme 
conditions of pressure and/or temperature are created in the interface of the colliding 
nuclei that simulate those existing in the interior of compact stellar objects or in the 
beginning of the universe. 
 In both terrestrial experiments and astrophysical observations, one of the main goals  
is the determination of the microscopic description of the EOS of dense nuclear matter in 
the framework of the fundamental theory of strong interactions (QCD).

  The current treatments have led to theoretical results which are unable to fully explain 
the observed phenomena of compact astrophysical objects.
  Among these treatments, the Nambu-Jona-Lasinio (NJL) model \cite{NJL1,NJL2} and MIT Bag model 
\cite{MIT} have been used in the study of compact stars to describe quark matter at nonzero 
temperature and density \cite{Koh,DPM1,DPM2}.
  However, in both the NJL and MIT Bag models, quarks enter in the respective EOS as a free quark 
gas with Fermi-Dirac distribution.
  In the EOS of the NJL model, the confinement is not explicitly included, whereas in the MIT 
Bag model the confinement is represented by a bag enclosing the free quarks with a constant 
$B$ which gives the energy density difference between the confined and deconfined phases.
  At low temperatures, as in compact star interiors, the NJL and MIT Bag EOS's are used 
in the large density region where the approximation of free quarks can be considered.  
  This is not so at all densities and (low) temperatures of quark matter.
  Quarks strongly interact submitted to a potential that is large when the ${\rm q\bar q}$ distances 
are large. 
  In this case, a nonperturbative method must be considered.

  Recently the EOS of quark-gluon plasma was derived in the framework of the Field Correlator 
Method (FCM) \cite{Si6}. 
 The FCM (for a review see \cite{DiG}) is a nonperturbative approach which naturally includes from 
first principles the dynamics of confinement in terms of color electric and color magnetic correlators. 
 The parameters of the model are the gluon condensate $G_2$ and the ${\rm q\bar q}$ interaction 
potential $V_1$ which govern the behavior of the EOS, at fixed quark masses and temperature.
 The FCM has been used to describe the deconfinement phase transition \cite{ST1,ST2,KS1,KS2} and 
an important feature of the model is that it covers the entire phase diagram plane, from the large 
temperature and small density region to the small temperature and large density region.
  In the connection between the FCM and lattice simulations, the critical temperature at 
$\mu_c=0$ turns out to be $T_c\sim170$ MeV for $G_2=0.00682\;{\rm GeV}^4$ \cite{ST1,ST2}.
 
 Very recently, the application of the FCM to the study of neutron stars (NS's) has been considered 
in Ref. \cite{Bur,Bal} where the microscopic EOS provided by the FCM is assumed for the internal 
NS cores.
  A farther interesting issue is the application of the FCM to the study of the strange quark 
matter (SQM) made of uds quarks and a class of compact stars  called {\it strange stars} (SS's).
  The existence of SQM  was first conjectured by Bodmer \cite{Bod} and Witten \cite{Wit}.
  Such a strange matter at zero temperature and pressure would have a ground state lower than 
that of the $^{56}{\rm F_e}$ nucleus and be possibly found in NS's \cite{Bod,Wit}. 
  The possibility of the SS's existence has been modeled in the framework of the MIT Bag model 
\cite{HZS,AFO}. 
  In contrast to this model, the FCM naturally includes the confinement which manifests itself in 
the gluon condensate and the ${\rm q\bar q}$ interaction potential.
  
  Observationally, there is growing evidences of existing stars made of SQM.  
  Among the possible SS's candidates is the X-ray pulsar Her X-1 discovered in 1972 
\cite{Tan}, the nearly isolated X-ray source RX J1856.35-3754 which has been studied in 
\cite{Dra}, the transient X-ray sources SAX J1808.4-3658 and XTE J1739-285 \cite{Kaa} 
and the atoll source 4U 1728-34. 
The general trend is that for these sources the SS hypothesis has produced mass-radius relations 
(MRR's) more consistent with the semi-empirical MRR's obtained from observations \cite{XDL1,XDL2} 
than those obtained from the NS hypothesis. 
 Very recently, the mass of the millisecond pulsar J1614-2230 has been measured using Shapiro 
delay technique \cite{Dem}.  By comparing the results among different EOS, the authors conclude 
that the high pulsar mass can only be interpreted in terms of a strongly interacting quark matter.

  In the present paper we made an application of the FCM to investigate the properties of the SS's 
in terms of the main parameters of the model. 
 Our aim is to understand the role of the nonperturbative dynamics of confinement in the 
description of the SQM with charge neutrality and $\beta$-equilibrium and the effects on the 
MRR's of SS's.

  This paper is organized as follows. In Sec. \ref{beqs}, we present the basic formalism of 
the FCM for the calculations of the important quantities.
 In Sec. \ref{res}, we show some properties of the strange matter in $\beta$-equilibrium and 
charge neutrality and calculate the strange star properties in terms of the model parameters.  
 In Sec. \ref{frem}, we summarize the main conclusions.

\section{Basic equations}
\lb{beqs}

 Let us summarize the main aspects of the FCM.
 The gluon field $A_\mu$ is assumed to be split into a nonperturbative 
background field $B_\mu$ and a (valence) perturbative quantum field $a_\mu$,
\bq
A_\mu=B_\mu+a_\mu~.
\lb{A}
\eq
 According to this description, the partition function of a system of quarks 
and gluons is given by
\bq
{\cal Z}^{SLA}_q=\bigg< N\int{\cal D}\phi\exp\bigg[-\int^{\beta=1/T}_0
d\tau\int d^3x\;L_{tot}({\bf x},\tau)\bigg]\bigg>_B
\lb{Z1}
\eq
where $\phi$ denotes all fields $a_\mu,\;\Psi,\;\Psi^+$ and $L_{tot}$ is the QCD 
Lagrangian density \cite{DiG}, and $N$ is a normalization constant. 
 In Eq.(\ref{Z1}), $<...>_B$ means averaging over nonperturbative background fields $B_\mu$.
\bigskip

 In the FCM approach, the confined-deconfined phase transition is dominated by the 
nonperturbative correlators. 
 The dynamic of deconfinement is described by Gaussian (quadratic in $F^a_{\mu\nu}F^a_{\mu\nu}$)  
colorelectric and colormagnetic gauge invariant  Fields Correlators $D^E(x)$, $D^E_1(x)$, 
$D^H(x)$, and $D^H_1(x)$. 
 The main quantity which governs the nonperturbative dynamics of deconfinement is given by the 
two point functions (after a decomposition is made) 
\bqn
g^2\bigg<\hat tr_f[E_i(x)\Phi(x,y)E_k(x)\Phi(y,x)]\bigg>_B&=&\delta_{ik}
[D^E+D^E_1+u^2_4\frac{\partial D^E_1}{\partial u^2_4}]+
u_iu_k\frac{\partial D^E_1}{\partial u^2}
\lb{gEE}
\eqn
\bqn
g^2\bigg<\hat tr_f[H_i(x)\Phi(x,y)H_k(x)\Phi(y,x)]\bigg>_B&=&\delta_{ik}
[D^H+D^H_1+u^2_4\frac{\partial D^H_1}{\partial u^2_4}]-
u_iu_k\frac{\partial D^H_1}{\partial u^2}
\lb{gHH}
\eqn
where $u=x-y$ and 
\bq
\Phi(x,y)=P\exp\bigg[ig\int^y_xA_\mu dx^\mu\bigg]
\lb{29}
\eq
is the parallel transporter (Schwinger line) to assure gauge invariance.

  In the confined phase (below $T_c$), $D^E(x)$ is responsible for confinement with string 
tension $\sigma^E=(1/2)\int D^E(x)d^2x$. 
 Above $T_c$ (deconfined phase), $D^E(x)$ vanishes while $D^E_1(x)$ remains nonzero being 
responsible (toghether with the magnetic part due to $D^H(x)$ and $D^H_1(x)$) for nonperturbative 
dynamics of the deconfined phase. 
  In lattice calculations, the nonperturbative part of $D^E_1(x)$ is parametrized in the form 
  \cite{Si5,DiG}
\bq
D^E_1(x)=D^E_1(0)\ex^{-|x|/\lambda}\;,
\lb{DE1}
\eq
where $\lambda=0.34\;{\rm fm}$ {\rm(full QCD)} is the correlation length, with the normalization 
fixed at $T=\mu=0$,
\bq
D^E(0)+D^E_1(0)=\frac{\pi^2}{18}G_2\;,
\lb{DEDE1}
\eq
where $G_2=0.012\pm0.006\;{\rm GeV}^4$ is the gluon condensate \cite{SVZ}.
 Another important quantity considered in the FCM is the large distance static ${\rm q\bar q}$ 
potential given by \cite{Si5,Si6,DiG}
\bq
V_1=\int^\beta_0d\tau(1-\tau T)\int^\infty_0\xi d\xi D^E_1(\sqrt{\xi^2+\tau^2})\;.
\lb{V1}
\eq
 
 The generalization of the FCM at finite $T$ and $\mu$ provides expressions  for the 
thermodynamics quantities where the leading contribution is given by the interaction of 
the single quark and gluon lines with the vacuum (called single line approximation (SLA)).  
 From Ref. \cite{Si6} and standard thermodynamical relations \cite{Kap}, we explicitly rewrite 
in more suitable forms to our calculations the equations (for one quark system, $N_f=1$) of 
the pressure 
\bq
p^{SLA}_q=\frac{1}{3}
\frac{2N_c}{(2\pi)^3}\int d^3k\frac{k^2}{E}
\bigg[f^{SLA}_q(T,J^E,\mu_q)+{\bar f}^{SLA}_q(T,J^E,\mu_q)\bigg]\;,
\lb{p}
\eq
the energy density
\bq
\varepsilon^{SLA}_q=
\frac{2N_c}{(2\pi)^3}\int d^3k´
\bigg[E-T(T\frac{\partial J^E}{\partial T}+\mu_q\frac{\partial J^E}{\partial\mu_q})\bigg]
\bigg[f^{SLA}_q(T,J^E,\mu_q)+{\bar f}^{SLA}_q(T,J^E,\mu_q)\bigg]\;,
\lb{e}
\eq
and include the number density of the quark system
\bqn
n^{SLA}_q&=&
\frac{2N_c}{(2\pi)^3}\int d^3k 
\bigg[f^{SLA}_q(T,J^E,\mu_q)-{\bar f}^{SLA}_q(T,J^E,\mu_q)\bigg]\nonumber\\
&-&T\frac{\partial J^E}{\partial\mu_q}\frac{2N_c}{(2\pi)^3}\int d^3k 
\bigg[f^{SLA}_q(T,J^E,\mu_q)+{\bar f}^{SLA}_q(T,J^E,\mu_q)\bigg]\;,
\lb{dens}
\eqn
where
\bqn
f^{SLA}_q(T,\mu_q,J^E)=\frac{1}{\ex^{\beta(E+TJ^E-\mu_q)}+1}\;\;&{\rm and}&\;\;
{\bar f}^{SLA}_q(T,\mu_q,J^E)=\frac{1}{\ex^{\beta(E+TJ^E+\mu_q)}+1}\;,
\lb{nJE}
\eqn
$E=\sqrt{k^2+m^2_q}$, $\beta=1/T$, and $J^E\equiv V_1/2T$ is the Polyakov loop exponent.  
 In order to give Eqs. (\ref{p})-(\ref{dens}) in its most general forms, we assume that 
$V_1$ is, in principle, a function of temperature and chemical potential. 
 However, according to the parametrization given by Eq. (\ref{V1}),  $V_1$ does not depend 
on the chemical potential.  
 As pointed out in \cite{ST1}, the expected $\mu$-dependence of $V_1$ should be weak 
for values of $\mu$ much smaller than the scale of vacuum fields (which is of the order 
of $\sim 1.5$ GeV) and is partially supported by the lattice simulations \cite{Dor}.
 As in \cite{Bur,Bal}, in this work we take $V_1$ constant. 

 The pressure and energy density of gluons are given by
\bq
p^{SLA}_{gl}=\frac{(N_c^2-1)}{3}\frac{2}{(2\pi)^3}\int d^3k
\frac{k}{\ex^{\beta(k+T{\tilde J}^E)}-1}
\lb{pgl}
\eq
and
\bq
\varepsilon^{SLA}_{gl}=3\;p_{gl}-T^2\frac{\partial\tilde{J}^E }{\partial T}(N_c^2-1)
\frac{2}{(2\pi)^3}\int d^3k\frac{1}{\ex^{\beta(k+T{\tilde J}^E)}-1}\;,
\lb{egl}
\eq
where ${\tilde J}^E=\frac{9}{4}J^E$ is the Polyakov loop exponent in the adjoint representation.

 In order to take into account the presence of electrons to keep the quark matter in 
$\beta$-equilibrium and with charge neutrality, we also include the equations for the pressure, 
energy density and number density of electrons given by
\bq
p_{\rm e}=\frac{1}{3}\frac{2}{(2\pi)^3}\int d^3k\frac{k^2}{E_\ex}
[f_{\rm e}(T,\mu_{\rm e})+{\bar f}_{\rm e}(T,\mu_{\rm e})]\;,
\lb{pel}
\eq
\bq
\varepsilon_{\rm e}=\frac{2}{(2\pi)^3}\int d^3k\;E_\ex
[f_{\rm e}(T,\mu_{\rm e})+{\bar f}_{\rm e}(T,\mu_{\rm e})]\;
\lb{eel}
\eq
and
\bq
n_{\rm e}=\frac{2}{(2\pi)^3}\int d^3k
[f_{\rm e}(T,\mu_{\rm e})-{\bar f}_{\rm e}(T,\mu_{\rm e})]\;,
\lb{nel}
\eq
where
\bqn
f_{\rm e}(T,\mu_{\rm e})=\frac{1}{\ex^{\beta(E_\ex-\mu_{\rm e})}+1}\;\;&,&\;\;
{\bar f}_{\rm e}(T,\mu_{\rm e})=\frac{1}{\ex^{\beta(E_\ex+\mu_\ex)}+1}\;
\lb{fdel}
\eqn
and $E_\ex=\sqrt{k^2_\ex+m^2_\ex}$ .

 Inside a SS, the composition of quark matter at a given baryon density or 
chemical potential must be in equilibrium with respect to weak interactions and the 
overall charge neutrality must be maintained. 
 The weak interactions reactions are given by
\bq
d\rightarrow u+\ex+{\bar\nu}_\ex
\lb{due}
\eq
and
\bq
s\rightarrow u+\ex+{\bar\nu}_\ex\;.
\lb{sue}
\eq
 The above reactions result in an energy loss by the star due to neutrino diffusion.  
 In this case, neutrino chemical potential may be set equal to zero and the chemical 
equilibrium is given by 
\bq
\mu_d=\mu_u+\mu_\ex\;
\lb{mud}
\eq
and
\bq
\mu_s=\mu_d\;.
\lb{mus}
\eq
 The overall charge neutrality requires that
\bq
\frac{1}{3}(2n^{SLA}_u-n^{SLA}_d-n^{SLA}_s)-n_\ex=0
\lb{chn}
\eq
where $n_i$ is the number density of particle $i$. By numerically solving 
Eqs. (\ref{mud})-(\ref{chn}), for each value of the input total number density 
\bq
n=n^{SLA}_u+n^{SLA}_d+n^{SLA}_s+n_\ex\;,
\lb{nT}
\eq
the unknown chemical potentials $\mu_{\rm u}$, $\mu_{\rm d}$, $\mu_{\rm s}$ and $\mu_{\rm e}$ 
are determined for fixed values of $T$, $G_2$ and $V_1$. 
 We here also include two useful quantities to express some of our results, namely, the 
baryon number density $n_B=(n^{SLA}_u+n^{SLA}_d+n^{SLA}_s)/3$ (also used in Eq. (\ref{ma})) 
and the baryon chemical potential $\mu_B=(\mu_u+\mu_d+\mu_s)/3$.

 The total pressure and energy density of the quark-gluon system, including the electrons 
inside the star, are given by
\bq
p=p^{SLA}_{gl}+\sum_{q=u,d,s}p^{SLA}_{q}-\Delta|\varepsilon_{vac}|+p_{\rm e}\;,
\lb{pqgl}
\eq
and
\bq
\varepsilon=\varepsilon^{SLA}_{gl}+\sum_{q=u,d,s}\varepsilon^{SLA}_{q}+
\Delta|\varepsilon_{vac}|+\varepsilon_{\rm e}\;,
\lb{eqgl}
\eq
where \cite{ST1,ST2}
\bq
\Delta|\varepsilon_{vac}|=\frac{11-\frac{2}{3}N_f}{32}\Delta G_2\;,
\lb{dvac}
\eq
is the vacuum energy density difference between confined and deconfined phases in terms of 
the respective difference between the values of the gluon condensate, 
$\Delta G_2=G_2(T<T_c)-G_2(T>T_c)\simeq \frac{1}{2}G_2$ and $N_f$ is the number of flavors.

  We use the quark masses $m_u=5$ MeV, $m_d=7$ MeV, $m_s=150$ MeV, the nuclear saturation 
density $n_0=0.153\;{\rm fm}^{-3}$, and the corresponding nuclear energy density
$\varepsilon_0=0.141\;{\rm GeV}{\rm fm}^{-3}$. 
  With exception of the phase diagram calculation, our investigation in the present work 
is made for T = 0 and $V_1$ constant. 
 Then, in Eqs. (\ref{e}), (\ref{dens}) and (\ref{egl}) we have $T^2\partial J^E/\partial T=-V_1/2$, 
$\;\partial J^E/\partial\mu_q=0\;$ and $\;T^2\partial{\tilde J^E}/\partial T=-9V_1/8$, respectively. 

 The masses and radii of sequences of strange stars are calculated by numerical integration 
of the Tolman-Oppenheimer-Volkov \cite{ShT} equilibrium equations for the mass and pressure 
given by
\bq
\frac{d}{dr}M(r)=4\pi^2\varepsilon(r)
\lb{dmdr}
\eq
and
\bq
\frac{d}{dr}p(r)=-G\;\frac{M(r)\varepsilon(r)}{r^2}\;
\frac{[1+p(r)/\varepsilon(r)][1+4\pi r^3p(r)/M(r)]}
{1-2GM(r)/r}\;,
\lb{tov}
\eq
where $\varepsilon$ is the total energy density and $G$ is the gravitational constant.
  For a given central value of $\varepsilon$, the numerical integration gives 
the gravitational mass $M$ and the stellar radius $R$.

  In our calculation, stability of the stars of a given sequence against disassembly of the 
equivalent number of neutrons to infinity is taken into account by comparing the gravitational mass
of each star calculated from Eq.(\ref{dmdr}) with the baryon mass $M_A$ given by \cite{Gle} 
(see also \cite{ZeN} for details)
\bq
M_A=4\pi\;m_n\int^R_0 n_B(r)[1-2GM(r)/r]^{-1/2}r^2dr\;.
\lb{ma}
\eq
 The stability against dispersion to infinity is always fulfilled along the sequence if $M<M_A$.
 Points for which $M\geq M_A$ in the star sequence are excluded.

\section{Results}
\lb{res}

 We concentrate our results on the main features of the application of the FCM to the 
calculation of SS properties.  
 We show some aspects of SQM with charge neutrality and in $\beta$-equilibrium  and 
investigate the SS properties in terms of the model parameters.

\subsection{Electrically neutral strange matter in $\beta$-equilibrium}
\lb{resA}

\begin{figure*}[th]
\centerline{
\psfig{figure=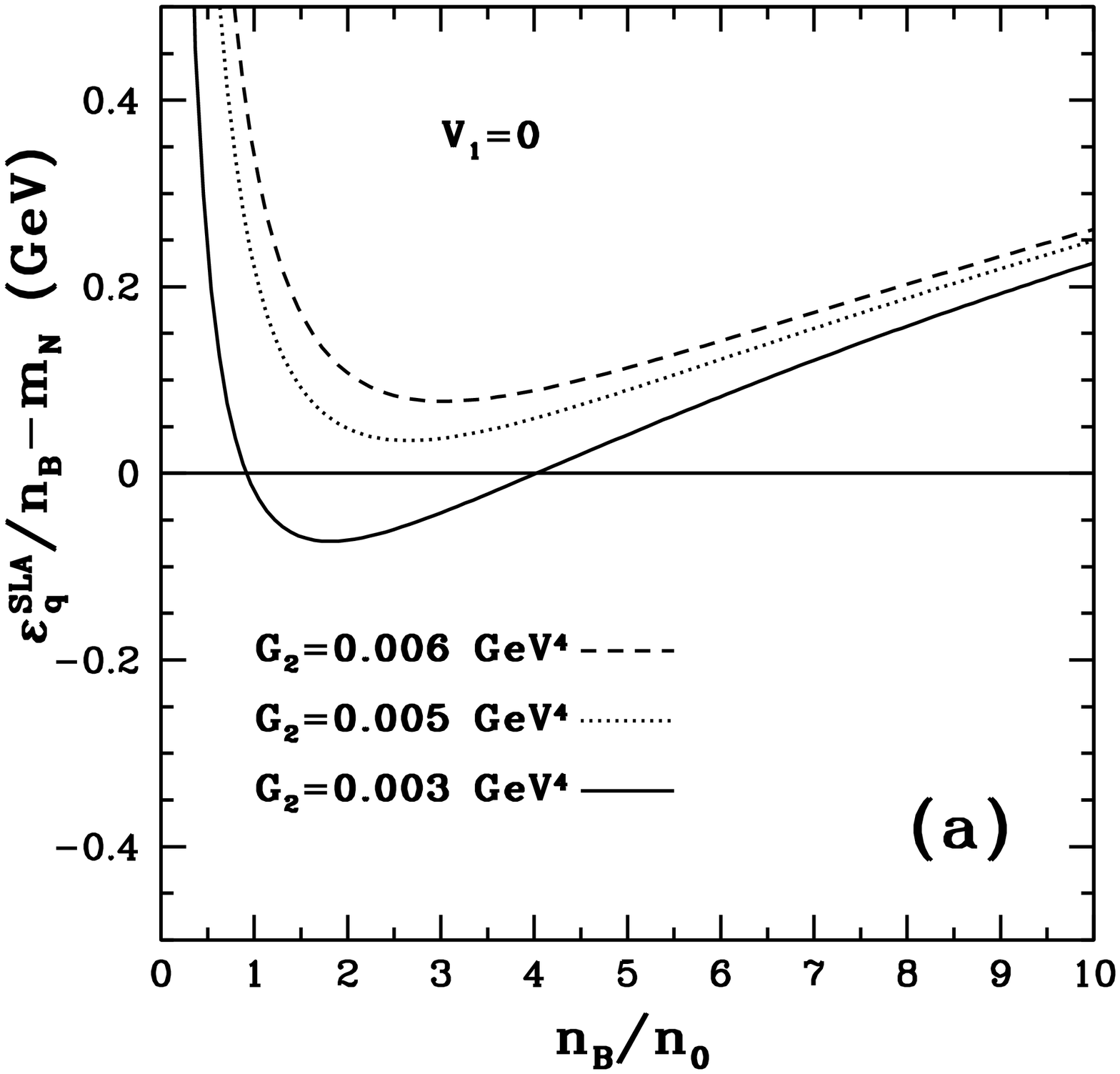,width=3.2truein,height=3.2truein}
\psfig{figure=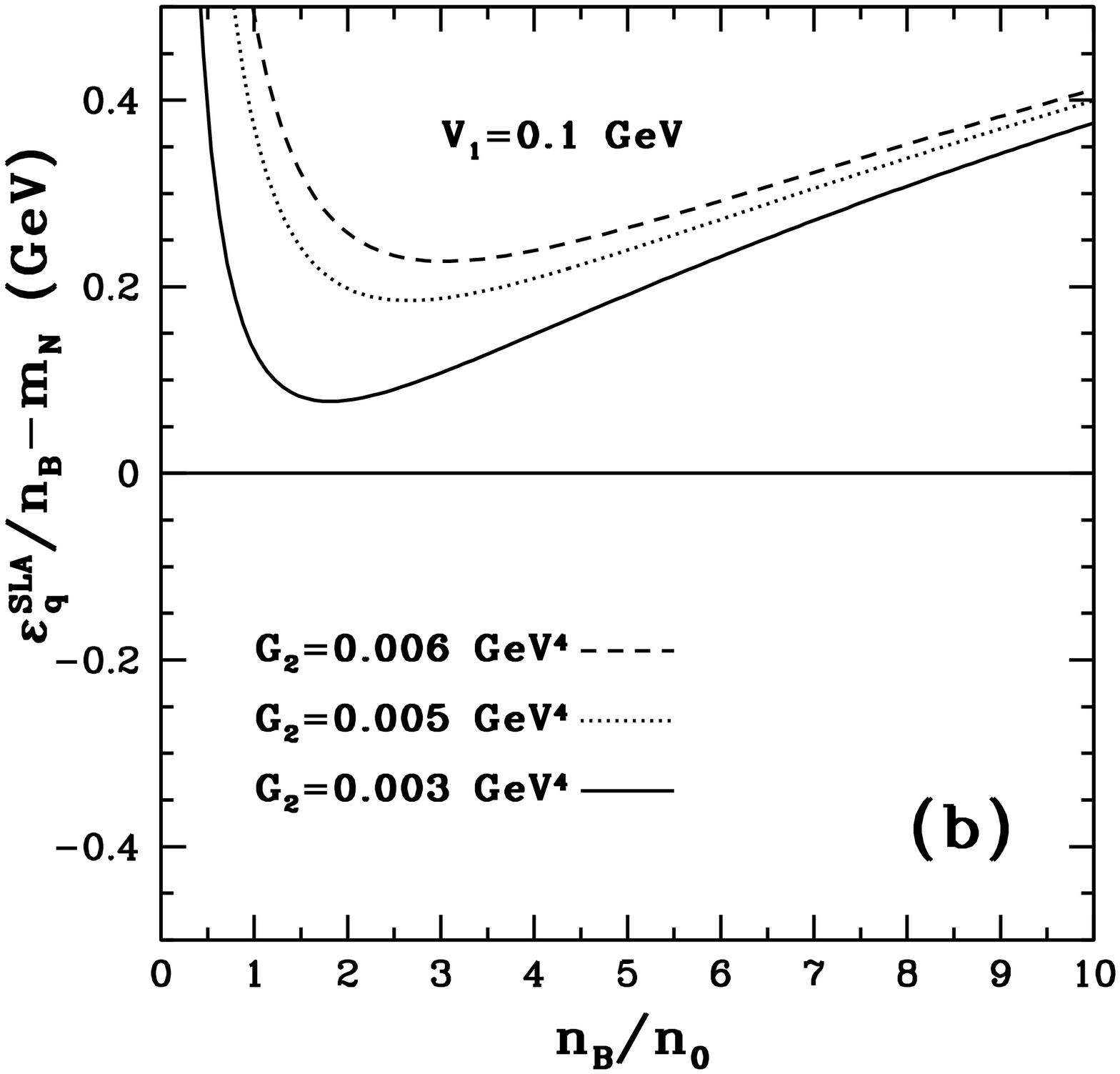,width=3.2truein,height=3.2truein}
\hskip .5in}
\caption{ 
Energy per baryon of the electrically neutral strange matter in $\beta$-equilibrium 
(minus the nucleon rest mass $m_N$), calculated for different choices of $G_2$
and $V_1$ at fixed temperature $T=0$, as function of the baryon number density 
$n_B$ normalized with respect to the nuclear number saturation density $n_0$.}
\label{eb0}
\end{figure*}

  In the framework of the FCM, the properties of the electrically neutral SQM 
in $\beta$-equilibrium are determined by the model parameters $G_2$ and $V_1$ . 
 In Fig. \ref{eb0} we plot the energy per baryon minus the nucleon mass as function of $n_B/n_0$ 
for different values of $G_2$ and $V_1$. 
 We note that $\varepsilon^{SLA}_q/n_B-m_N$ becomes larger as $G_2$ increases. 
 We also note that $\varepsilon^{SLA}_q/n_B-m_N$ increases as $V_1$ increases because 
$\mu_q$ ($q=u,d,s$) behaves as an increasing function of $V_1$ at fixed $T=0$ and 
$n_q^{SLA}$. 
 Such a feature which is expressed in Eqs. (\ref{e})-(\ref{nJE}) can be explored 
by using the values of $G_2$ and $V_1$ to verify the Bodmer-Witten conjecture \cite{Bod,Wit}.
 We come back to this subject in Sec. \ref{witt}.

\begin{figure*}[th]
\centerline{
\psfig{figure=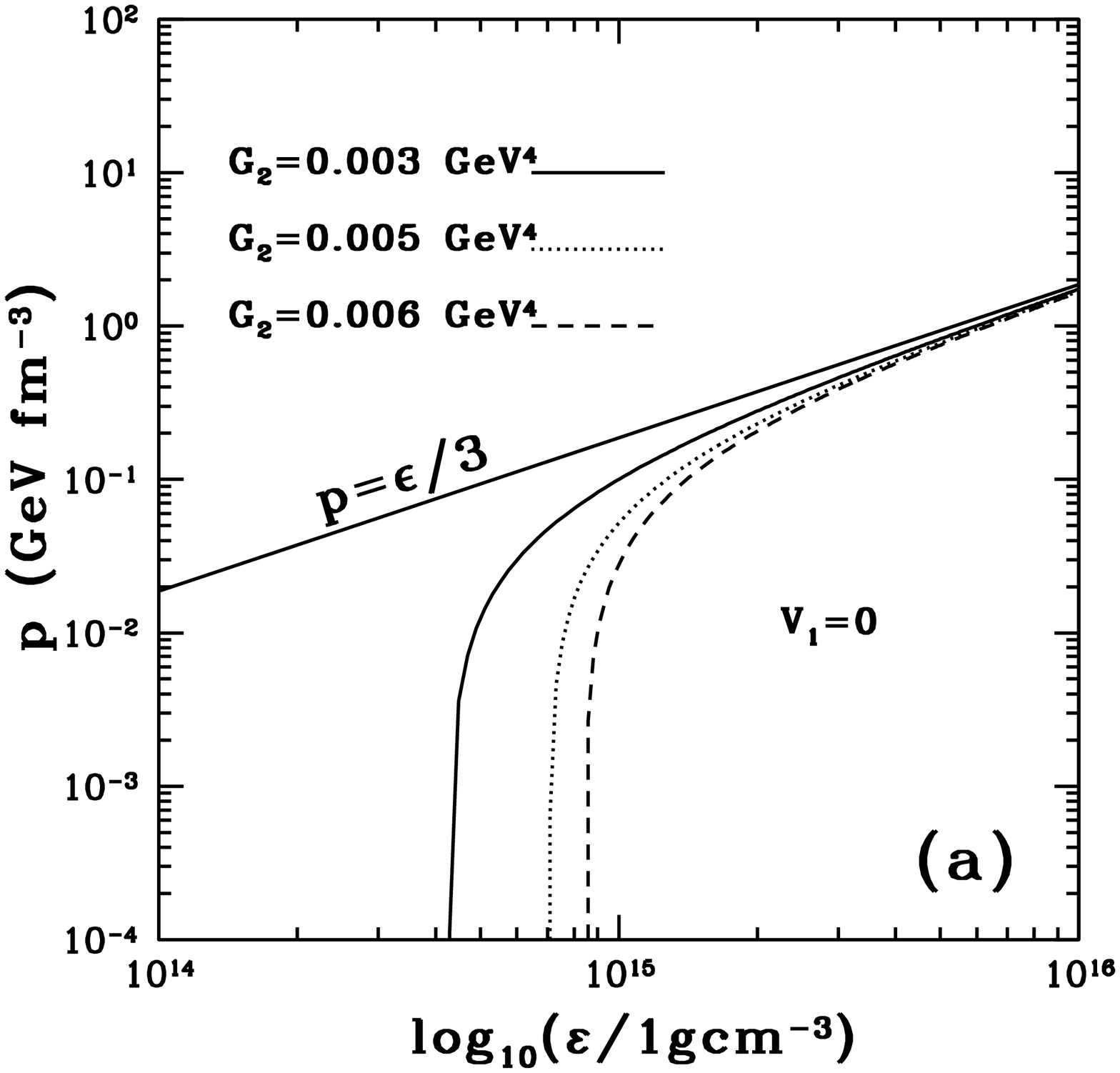,width=3.2truein,height=3.2truein}
\psfig{figure=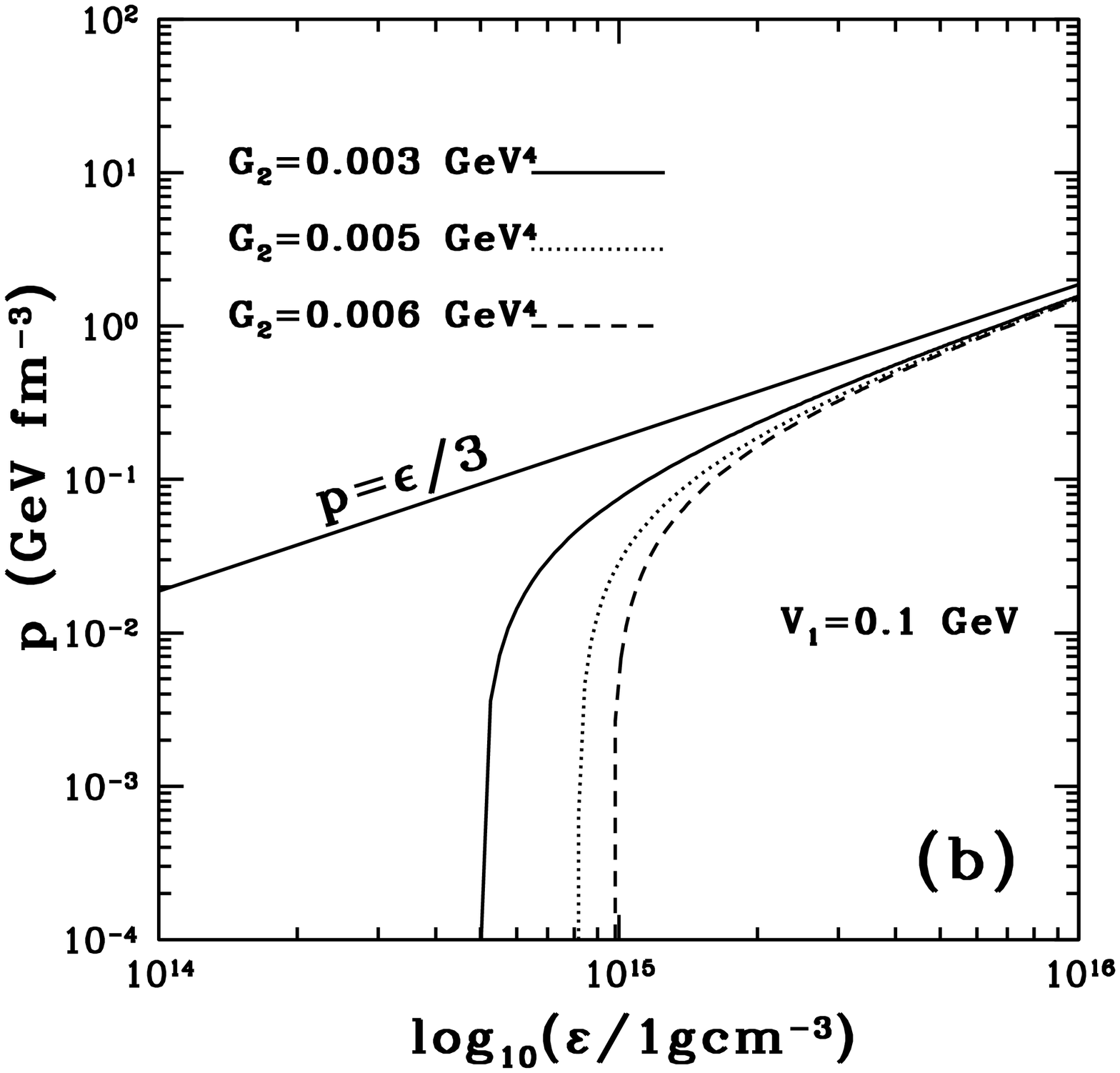,width=3.2truein,height=3.2truein}
\hskip .5in}
\caption{
Pressure of the electrically neutral strange matter in $\beta$-equilibrium, 
calculated for different choices $G_2$ and $V_1$ at fixed temperature $T=0$,  
as function of the energy density. 
The equation of state limit of a massless ideal gas ($p=\epsilon/3$) is also shown.}
\label{ep0a}
\end{figure*}

 Fig. \ref{ep0a} shows the sensitivity of the EOS  to the values of $G_2$ and/or $V_1$  
where the pressure is depicted as function of the energy density. 
 In this figure, we see  that the EOS becomes softer as $G_2$ increases.  
 From panel (a) to (b) we also observe a slight softening of the EOS when $V_1$ becomes larger. 
 These features are explicitly expressed in Eqs. (\ref{p}), (\ref{e}), and (\ref{nJE}) 
and represent the overall nonperturbative effects of color confining forces. 
 At this point it is instructive to calculate the speed of sound $v_s/c=(dp/d\varepsilon)^{1/2}$. 
For the values of $G_2$ and $V_1$ considered in this work we have obtained $v_s/c\;\gappl0.577$,  
the upper bound corresponding to $p=\varepsilon/3$.

 To proceed our investigation, some remarks concerning the calculation of SS masses must be made.  
 In this work, we are considering pure quark stars made of quarks u, d, s.  
 So, in order to avoid the quark-hadron phase transition when the density of the SS decreases 
from its maximum at the center of the star to its minimum at the star surface, we must chose the 
appropriate values of $G_2$ and $V_1$ at a given temperature.  
 In other words, $G_2$ and $V_1$ must be constrained by density values for which quark matter 
is deconfined and by allowed astrophysical possibilities for masses and radii. 
 To this end, we first obtain the phase transition curve $T_c(\mu)$ by using the condition 
$p(T_c)=p_H(T_c)$, where $p_H$ is the pressure of the confined (hadronic) phase.    
 Quark stars with crust have been currently considered in the literature. 
 Probably, in our calculations, a hadronic crust would affect the values of the parameters 
$G_2$ and $V_1$. 
 As long as the author knows, up to date a hadronic EOS has not been derived in the framework 
of the FCM as done for the quark-gluon plasma phase \cite{Si6}. 
 The inclusion of an EOS obtained from a  theory with diferent parameters would induce us to 
misunderstand the true role of $G_2$ and $V_1$. 
 In order to avoid wrong interpretation of the FCM parameters, we neglect the possible existence 
of a hadronic crust no matter how thin it may be.
 As done in \cite{ST1,ST2}, we work in the $p_H=0$ approximation.  
 Then, from Eq. (\ref{pqgl}) we obtain the phase diagram of the SQM with $\beta$-equilibrium 
and charge neutrality, by numerically solving the equation
\bq
\sum_{q=u,d,s}p^{SLA}_q+p^{SLA}_{qgl}-\Delta|\epsilon_{vac}|+p_{\rm e}=0\;,
\lb{phd}
\eq
together with Eqs. (\ref{mud})-(\ref{nT}), to  determine $\mu_u$, $\mu_d$, $\mu_s$, 
$\mu_{\rm e }$ and $T$ for given values of $n$, at fixed $G_2$ and $V_1$.

\begin{figure*}[th]
\centerline{
\psfig{figure=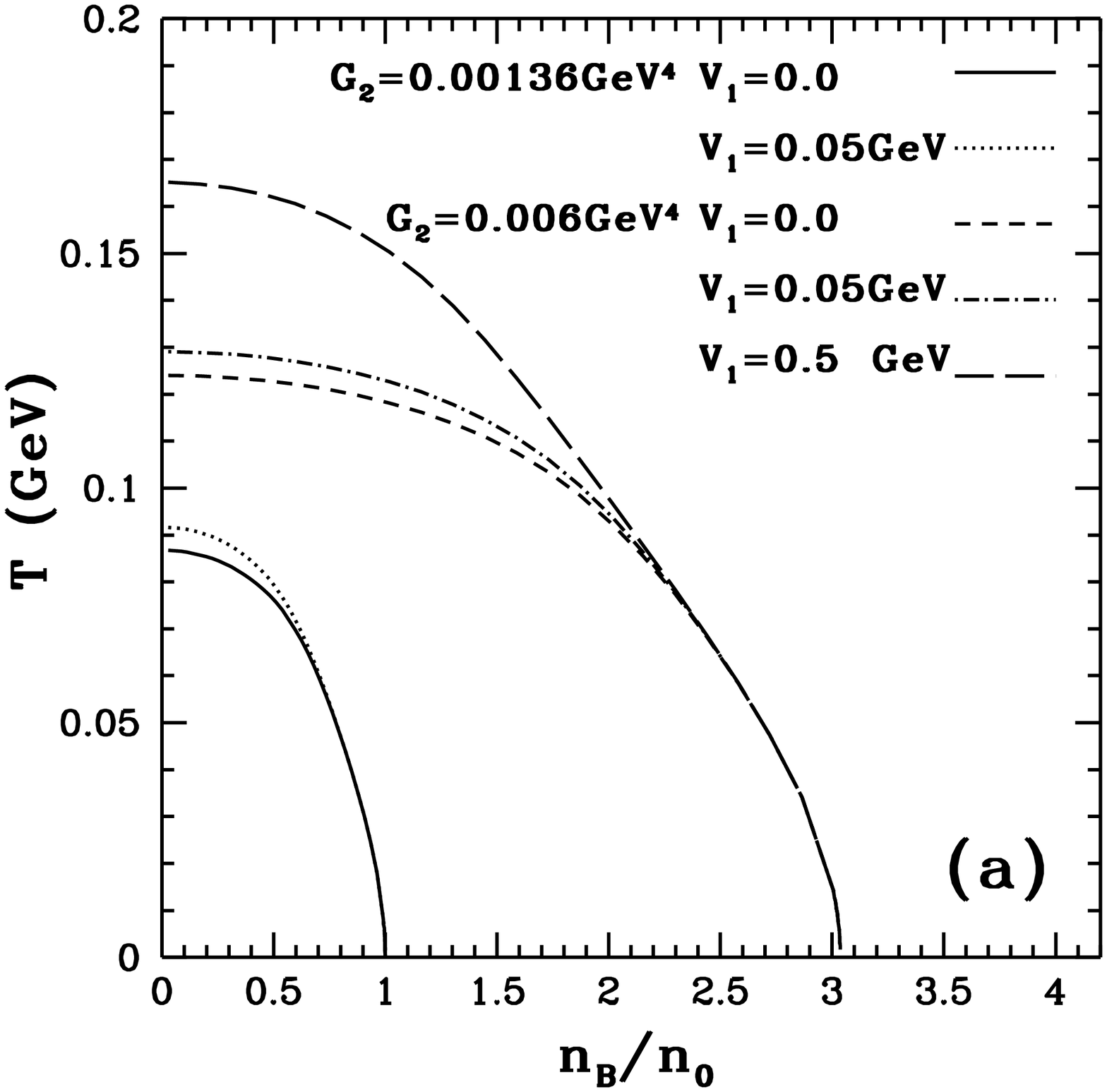,width=3.2truein,height=3.2truein}
\psfig{figure=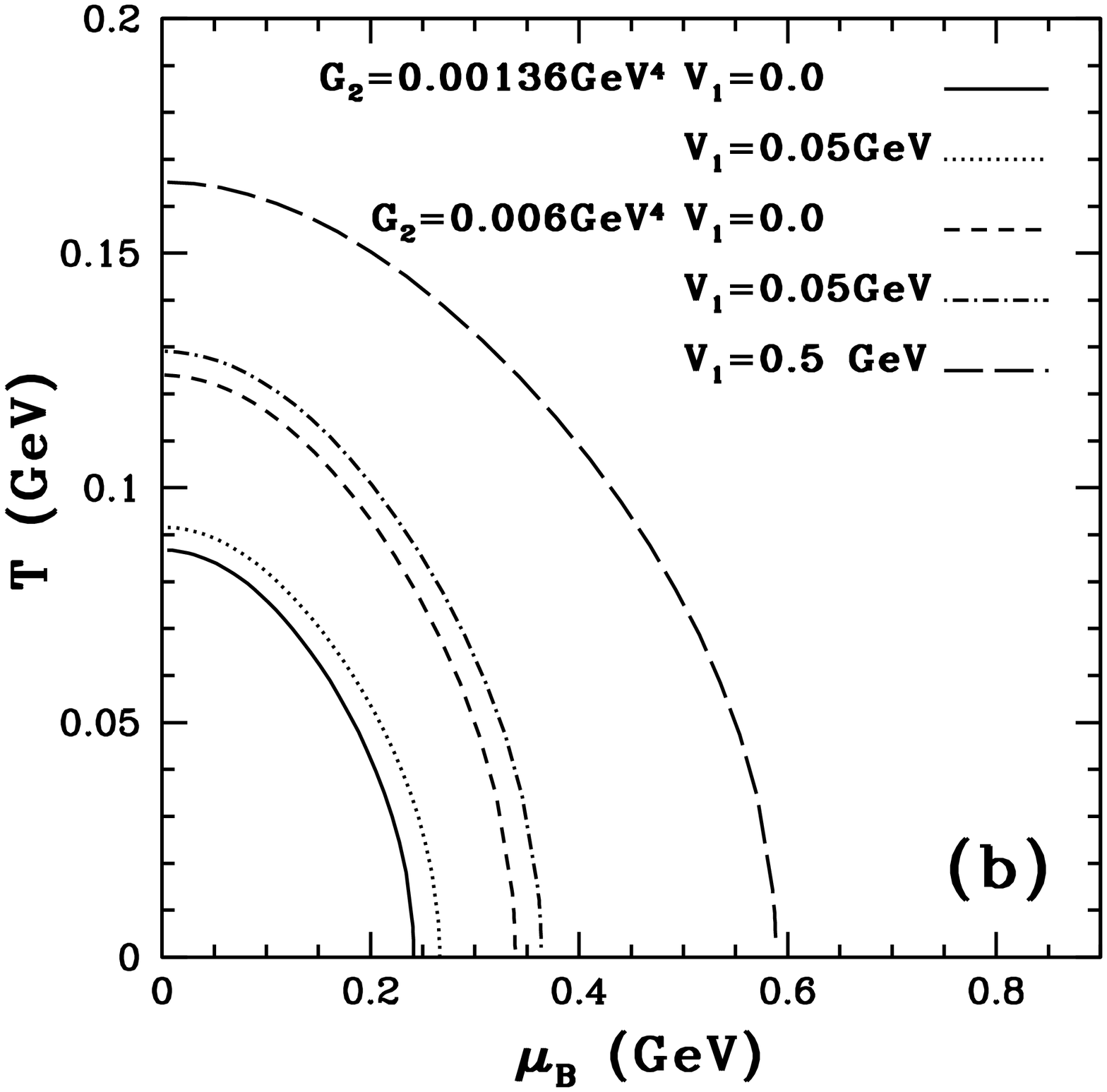,width=3.2truein,height=3.2truein}
\hskip .5in}
\caption{ 
Phase diagram of the electrically neutral strange matter in $\beta$-equilibrium 
calculated for different choices of $G_2$ and $V_1$. 
Panel (a): The temperature depicted as function of the baryon number density $n_B$ 
normalized with respect to the nuclear number saturation density $n_0$. 
Panel (b): The temperature depicted as function of the baryon chemical potential 
$\mu_B$.}
\label{tcrho}
\end{figure*}

 In Fig. \ref{tcrho}, $T$ is depicted for different values of $G_2$ and $V_1$ as function of 
$n_B/n_0$  in panel (a) and as function of $\mu_B$ in panel (b). 
 The general trend is that $T$ increases with the increase of $G_2$ and/or $V_1$ for fixed 
$n_B$($\mu_B$).
 In panel (a), we observe that at $T=0$ the curves corresponding to different values of $V_1$ 
converge to the same value of $n_B/n_0$ depending only on $G_2$ 
 \footnote{From Eq. (\ref{nJE}) we have $\mu_q=\sqrt{k^2+m_q^2}+V_1/2$ for $T\rightarrow0$. Thus, 
in Eqs. (\ref{mud}) and (\ref{mus}), $V_1/2$ is canceled (assuming $V_1$ independent of the 
chemical potentials of u,d,s quarks, of course).}.
 On the other hand, it is not exactly known the density at which quark-hadron phase transition 
takes place. 
 Due to this fact, we extrapolate our calculation to obtain $G_2=0.00136\;{\rm GeV}^4$ 
in order to give $n_B/n_0=1$ at $T=0$. 
 We use this value of $G_2$ as a reference one to guide us to examine the results for SS masses and radii .
 We anticipate that for $G_2>0.00136\;{\rm GeV}^4$ the results naturally show that $n_B/n_0>1$ 
inside the stars, as well as on its surfaces.
 Panel (b) shows the temperature as function of $\mu_B$ (for the same values of $G_2$ and $V_1$ of 
panel (a)). 
 For $V_1=0.5$ GeV and $G_2=0.006\;{\rm GeV}^4$, the phase transition curve $T_c(\mu)$ of the 
electrically neutral SQM in $\beta$-equilibrium is nearly the same that of \cite{ST1,ST2} where 
$T_c(0)\simeq0.170$ GeV for $n_f=3$. 
 The role of the electrons is to maintain the $\beta$-equilibrium and charge neutrality among the 
population of u,d,s quarks.
 Along the curve $T_c$, the electron number density is small, varying from $n_{\rm e}/n_B\sim10^{-2}$ 
at $\mu_B=0$ to $n_{\rm e}/n_B\gappl10^{-4}$ at $\mu_B\sim0.25-0.6$ GeV. 

\subsection{Strange star calculations}
\lb{ssc}

 Our next step is the calculation of gravitationally stable non-rotating SS sequences for 
several values of $G_2$ and $V_1$. 
 We divide our investigation in two steps. 
 First, independently of the results obtained in \cite{ST1,ST2} by comparison with lattice 
predictions, we construct a scenario for $G_2$ and $V_1$ to calculate the main properties of SS. 
 The calculation must here be taken only as an extrapolation beyond the results of \cite{ST1,ST2}  
to obtain values for $G_2$ and $V_1$ in order to calculate masses and radii of stable sequences.
 Second, we limit ourselves to the values of $G_2$ in \cite{ST1,ST2} as shown in Fig. \ref{stmr3a}. 

\begin{figure*}[th]
\centerline{
\psfig{figure=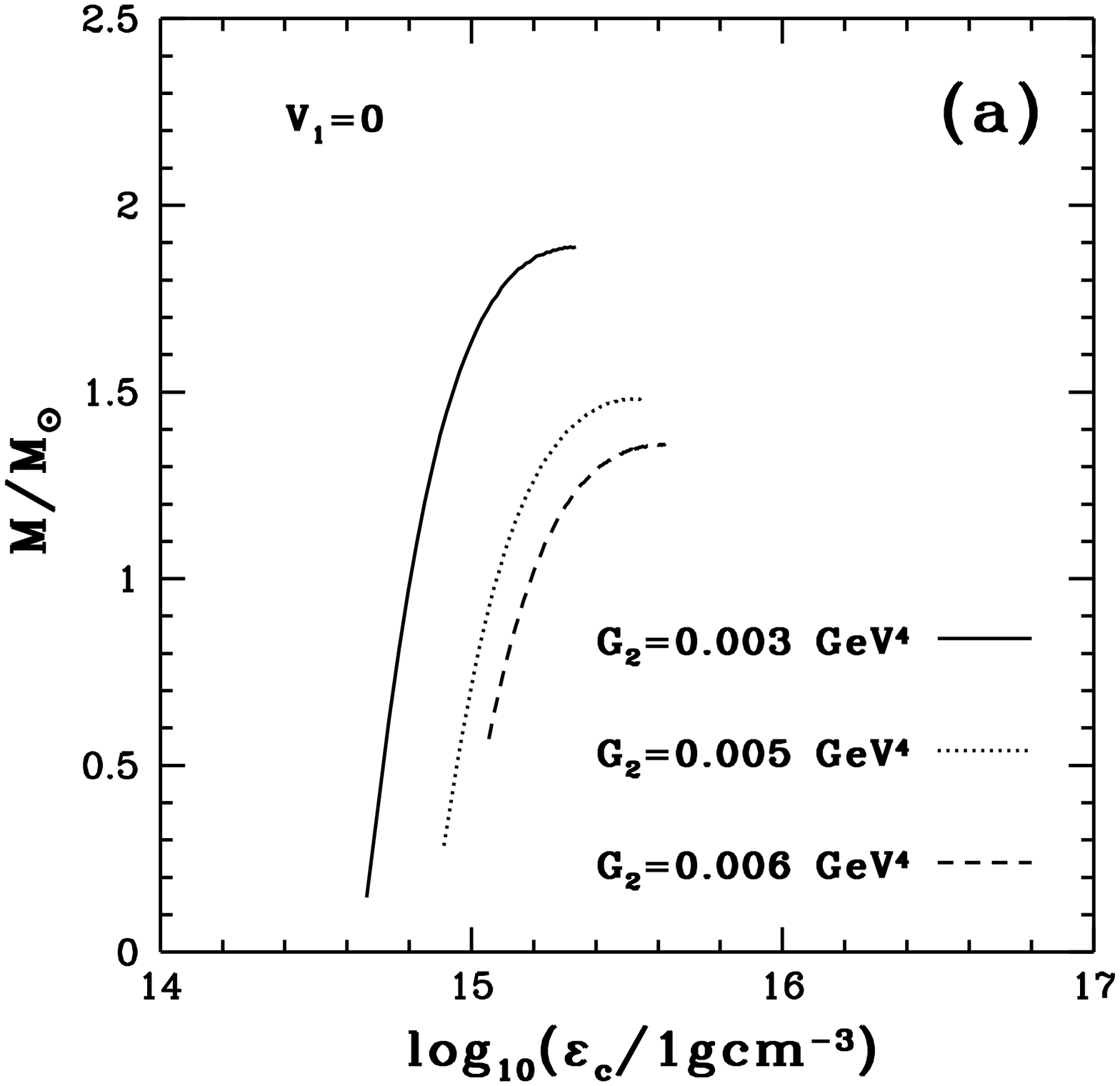,width=2.5truein,height=3.2truein}
\psfig{figure=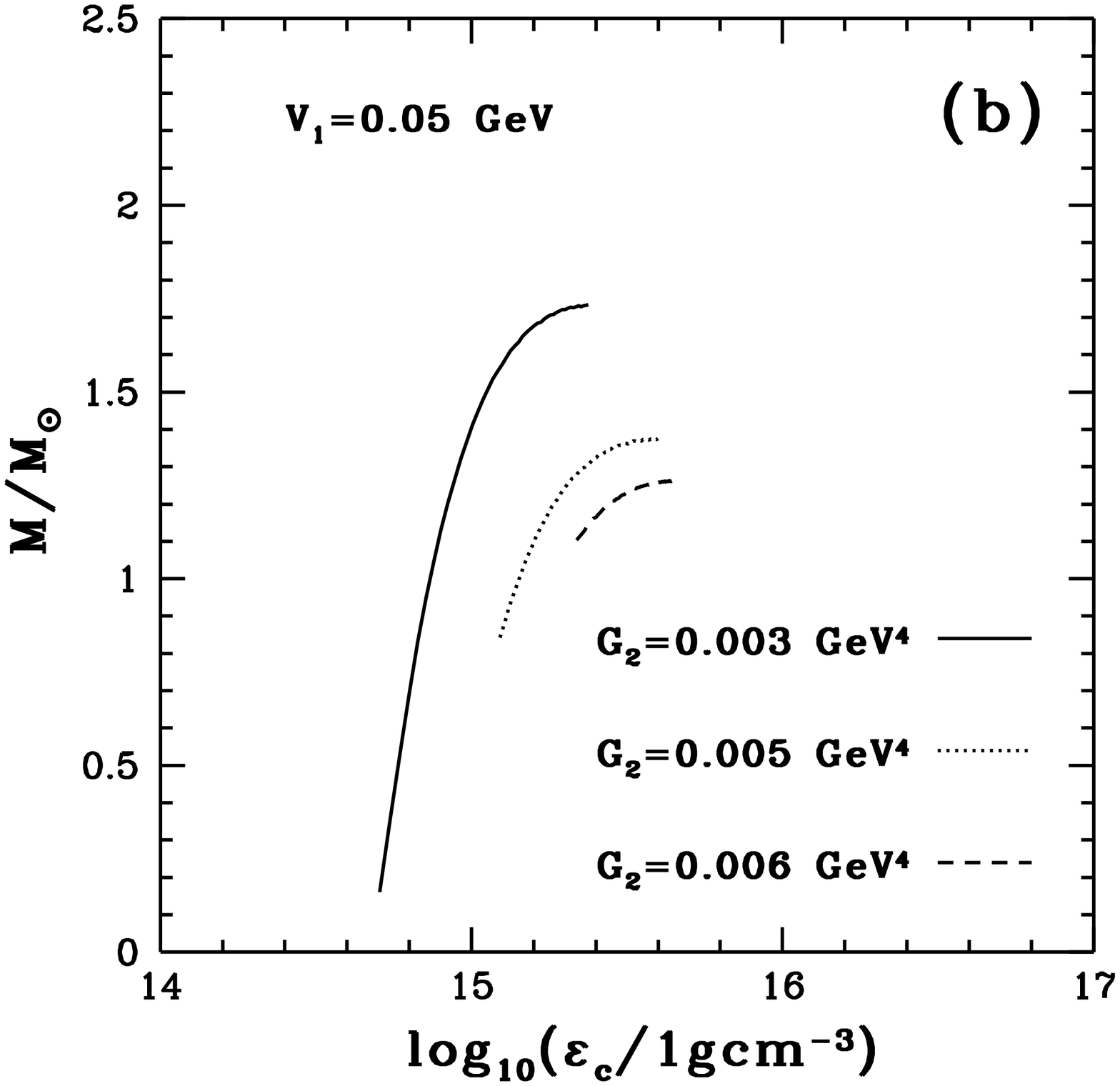,width=2.5truein,height=3.2truein}
\psfig{figure=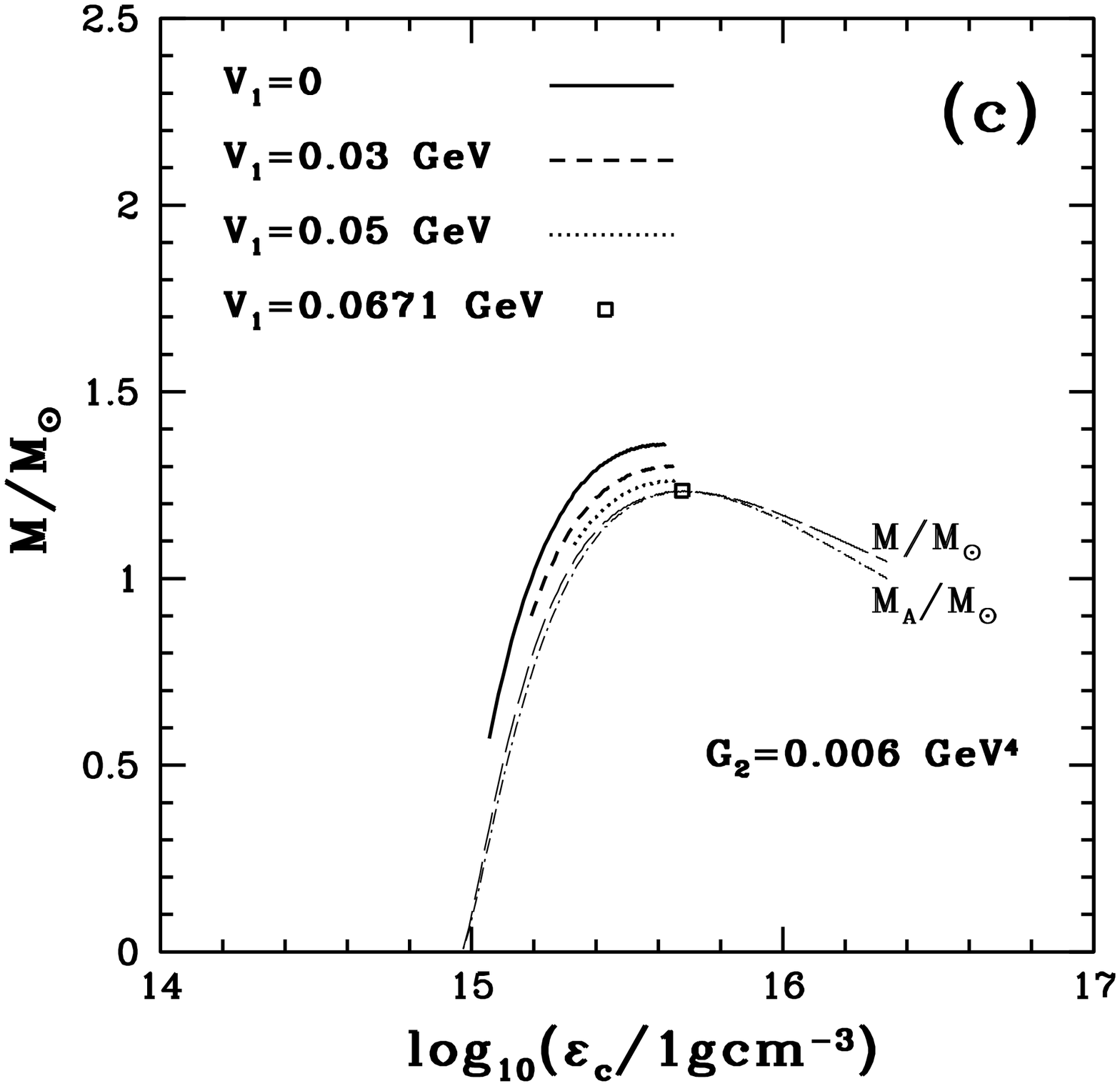,width=2.5truein,height=3.2truein}
\hskip .5in}
\caption{
The variation of $M/M_\odot$ calculated for different $G_2$ and $V_1$ at fixed 
temperature $T=0$. 
In panels (a) and (b), we see the shortening of the sequences when $G_2$ and/or $V_1$ 
become larger. 
Panel (c): For a fixed value of the gluon condensate, say $G_2=0.006\;{\rm GeV}^4$, 
the variation of the shortening of the sequences is shown for increasing values of $V_1$ 
up to the limit of only one star in the sequence for $V_1=0.0671\;{\rm GeV}$, 
indicated by the small open square dot. 
 The long-dashed ($M/M_\odot$) and dot-dashed ($M_A/M_\odot$) curves serve to guide the eye.}
\label{mst0a}
\end{figure*}

  We now calculate SS sequences at $T=0$ from Eqs. (\ref{dmdr}), (\ref{tov}) and (\ref{ma}) for 
fixed values of the FCM parameters.  
 In order to yield stable sequences, we constrain the values of $G_2$ and $V_1$ with respect to 
the condition $M<M_A$ and stability against gravitational collapse to a black hole. 
 The former is related to the binding energy of the stars and is responsible for the left end points 
of the sequences while the later determines the right end points at the maximum mass. 
 In Fig. \ref{mst0a} we plot the mass $M$ (in units of the solar mass $M_\odot=2\times10^{33}g$) 
as function of  the central energy density for different values of $G_2$ and/or $V_1$. 
 Each sequence begins at a minimum mass and ends at a maximum allowed mass; beyond this mass, 
the stars are no longer stable against gravitational collapse to a black hole.
 Between the left and right end points we have $M<M_A$ (the sequences for $M_A/M_\odot$ are 
not shown in panels (a) and (b)).

\begin{figure*}[th]
\centerline{
\psfig{figure=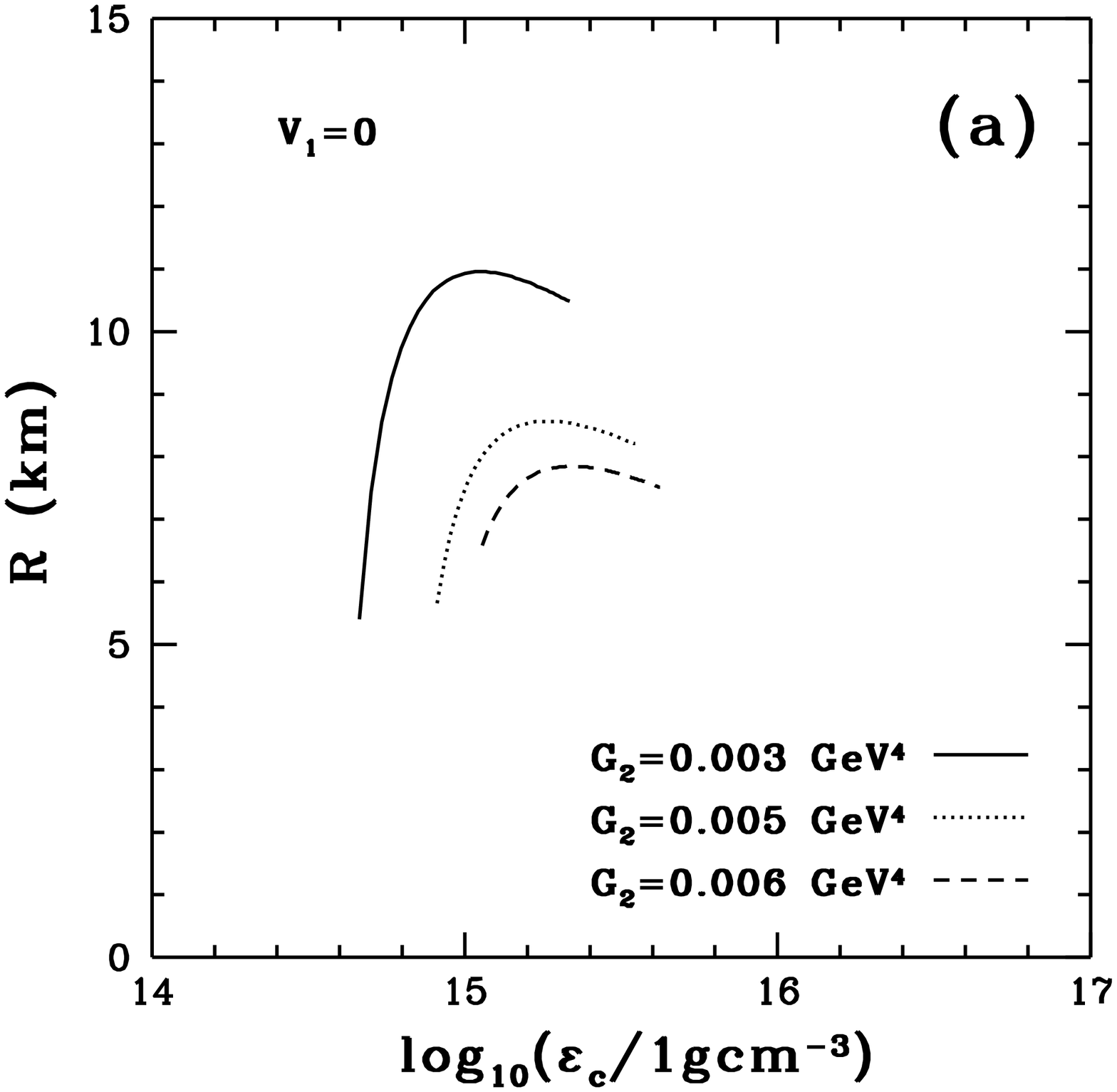,width=2.5truein,height=3.2truein}
\psfig{figure=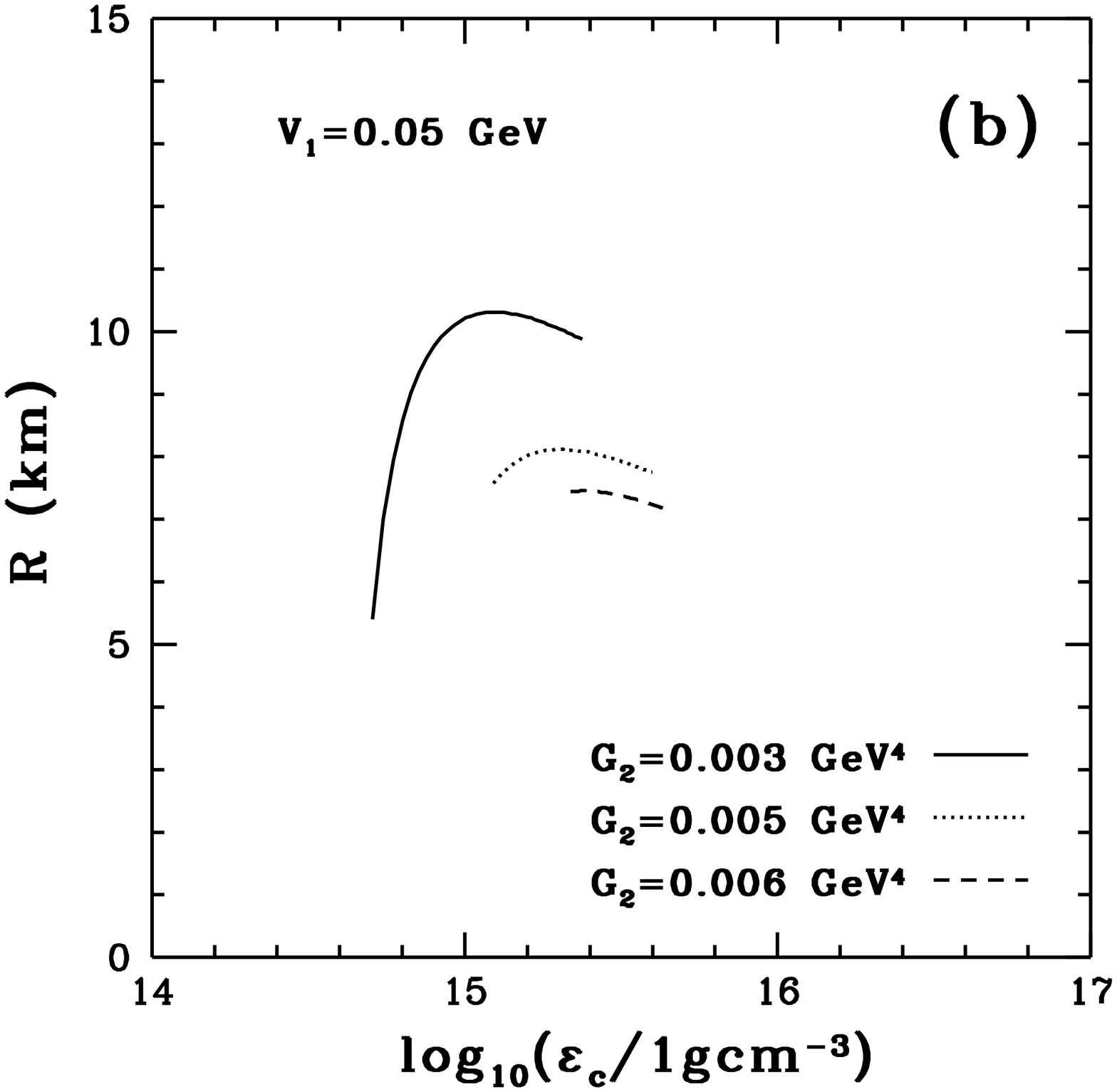,width=2.5truein,height=3.2truein}
\psfig{figure=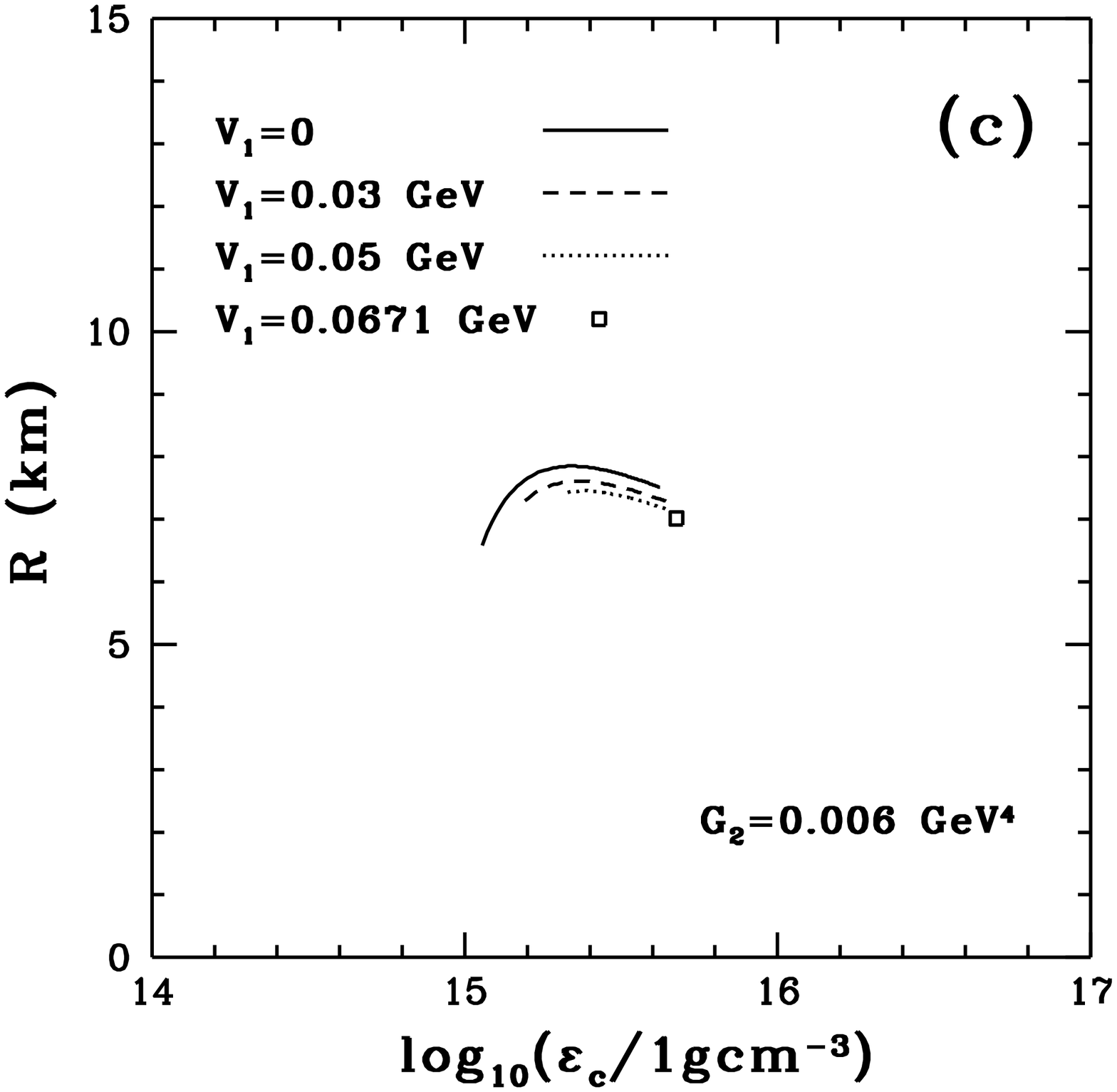,width=2.5truein,height=3.2truein}
\hskip .5in}
\caption{ 
The same as in Fig. \ref{mst0a}, but for the radius $R$.
In panel (c), the small open square dot  indicates the limiting sequence of 
only one star for $V_1=0.0671$ GeV.}
\label{rst0a}
\end{figure*}

\begin{figure*}[th]
\centerline{
\psfig{figure=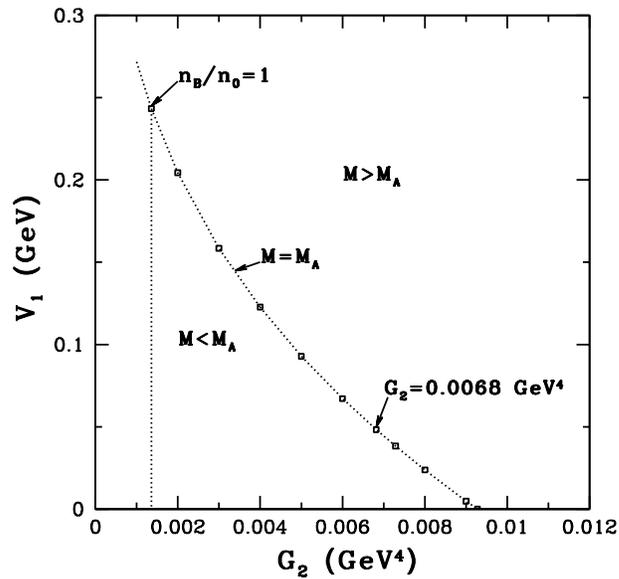,width=3.2truein,height=3.2truein}
\hskip .5in}
\caption{ 
Values of $G_2$ and $V_1$ for the limiting case of stable sequences 
with a single star at the lower maximum mass limit given by $M=M_A$ , 
as shown in panel (c) of Fig. \ref{mst0a} (small open square dot). 
Below the dotted curve all sequences are stable ($M<M_A$).}
\label{g2v1}
\end{figure*}

 In Fig. \ref{rst0a} we plot the respective radii, where the right end points correspond to 
the maximum masses shown in Fig. \ref{mst0a}.
 In both Figs. \ref{mst0a} and \ref{rst0a}, the masses and radii decrease as $G_2$ and/or $V_1$ 
increases because of the softening of the EOS. 
 It can also be noted that the lengths of the sequences are shortened, with the left end points 
going up to the respective right end peaks.
 We explore this behavior to determine $G_2$ and $V_1$ in order to obtain sequences in the 
limit of only one star at the maximum mass\footnote{The stars of an equilibrium configuration, 
obtained from Tolman-Oppenheimer-Volkov equations, pass from stability to instability (with 
respect to radial modes of oscillations) at a given central energy density 
$\varepsilon_c$ when the mass is stationary, e. g., $\partial M(\varepsilon_c)/\partial\varepsilon_c=0$. 
Moreover, the baryon number $N_A$ given by the integral in the r.h.s. of Eq. (\ref{ma}) 
is also stationary at the same value of $\varepsilon_c$ as the mass $M(\varepsilon_c)$ \cite{Gle}. 
 Then, it follows that $M(\varepsilon_c)$ and $M_A(\varepsilon_c)=m_nN_A(\varepsilon_c)$ are 
maxima at the same energy density (for a detailed analysis of stability, see \cite{HTW}).}
$M=M_A$, as the one represented by the small open square dot 
(where the excluded parts of the sequences for which $M>M_A$ are depicted to guide the eye), 
shown in panel (c) of Fig. \ref{mst0a}.
 Then, for $G_2=0.006\;{\rm GeV}^4$ and $V_1=0.0671\;{\rm GeV}$ we obtain a lower bound for 
the maximum masses of the sequences obeying the $M<M_A$ condition.
 For $V_1>0.0671$GeV, we have sequences with $M>M_A$ (which have been discarded). 
 We remark once again that the maximum masses diminish with the increase of $G_2$ and/or 
$V_1$, satisfying the $M<M_A$ requirement up to the lower bound given by $M=M_A$.
 In doing this for several values of $G_2$ and $V_1$, we obtain a curve in the plane of the 
FCM parameters where each point ($G_2,V_1$) represents a SS sequence in the limit of a single 
star at $M=M_A$, as shown in Fig. \ref{g2v1}. 
 Below the $M=M_A$ curve, all masses in the SS sequences satisfy $M<M_A$; otherwise, all 
sequences correspond to $M\geq M_A$. 

 An important point to be observed along the $M=M_A$ curve is that the baryon number density on 
the SS surface increase with the increasing value of $G_2$. 
 In particular, we have found that $n_B/n_0=1$ for $G_2=0.00136\;{\rm GeV}^4$. 
 This value of $G_2$ is too low with respect to lattice predictions to be realistic. 
 Then, greater values of the gluon condensate favor SS's surface densities higher than the nuclear ones. 
 On the other hand, the values of $V_1$ become lower along the $M=M_A$ curve. 
 Also shown in Fig. \ref{g2v1} is the point at $G_2=0.00682\;{\rm GeV}^4$ 
( $\Delta G_2=0.00341\;{\rm GeV}^4$ in Refs. \cite{ST1,ST2}) for which SS sequences may be obtained 
for $0\leq V_1<0.0483$ GeV. 
 It is worthwhile emphasize that differently from the lattice prescription which gives 
$V_1\simeq0.5$ GeV \cite{ST1,ST2}, this value can not be achieved by the astrophysical calculations of 
the present work, except for unrealistic low values of $G_2$ with respect to the lattice results. 
 For instance, extrapolating our calculation for $G_2=0.0005$ we obtain $V_1\sim0.33$ GeV. 
 We note that for $0< G_2\leq 0.00924\;{\rm GeV}^4$ astrophysical calculations constrain the 
values of $V_1$ to be $0\leq V_1<V_1(M=M_A)$ . 

\begin{figure*}[th]
\centerline{
\psfig{figure=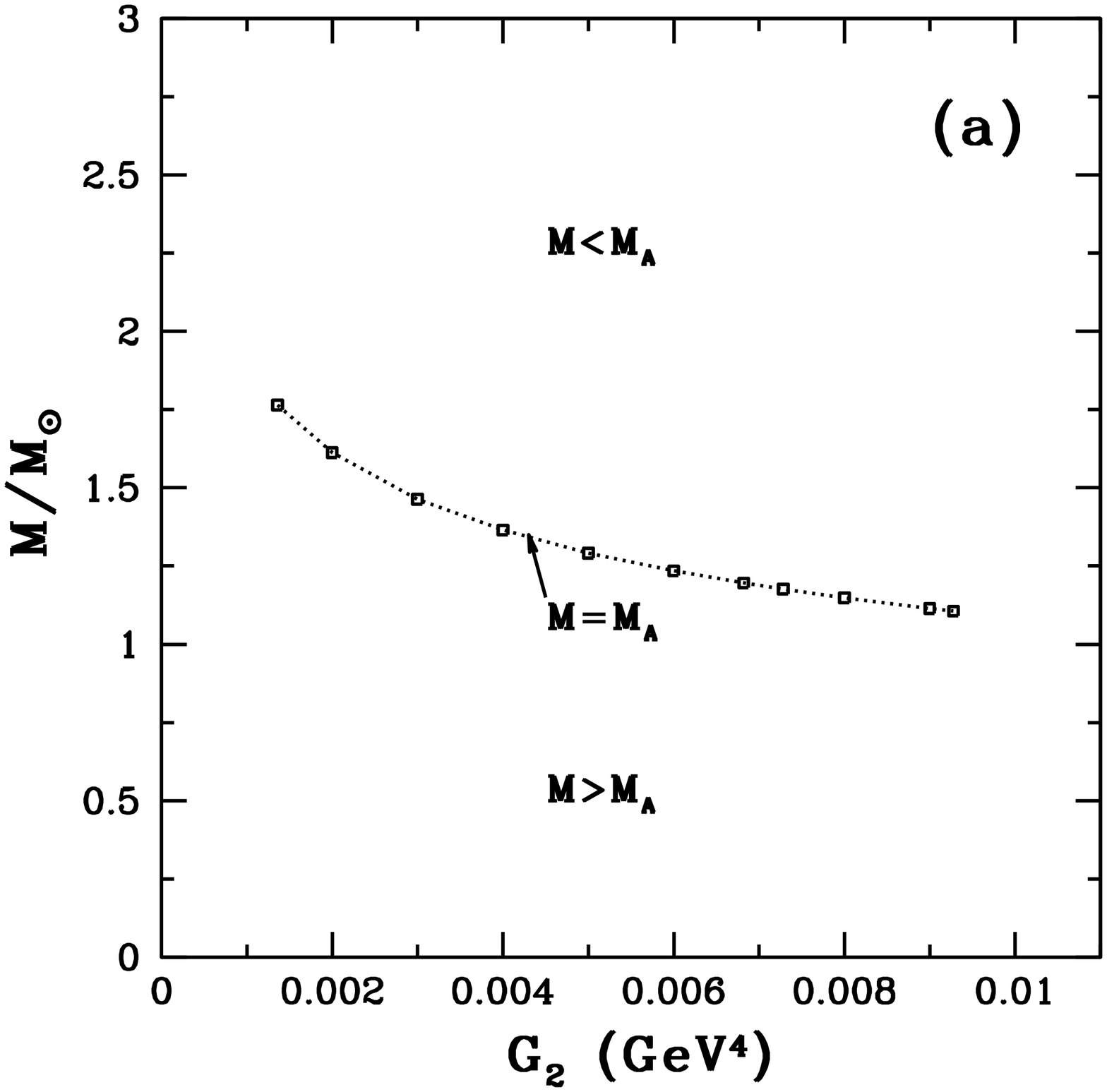,width=3.2truein,height=3.2truein}
\psfig{figure=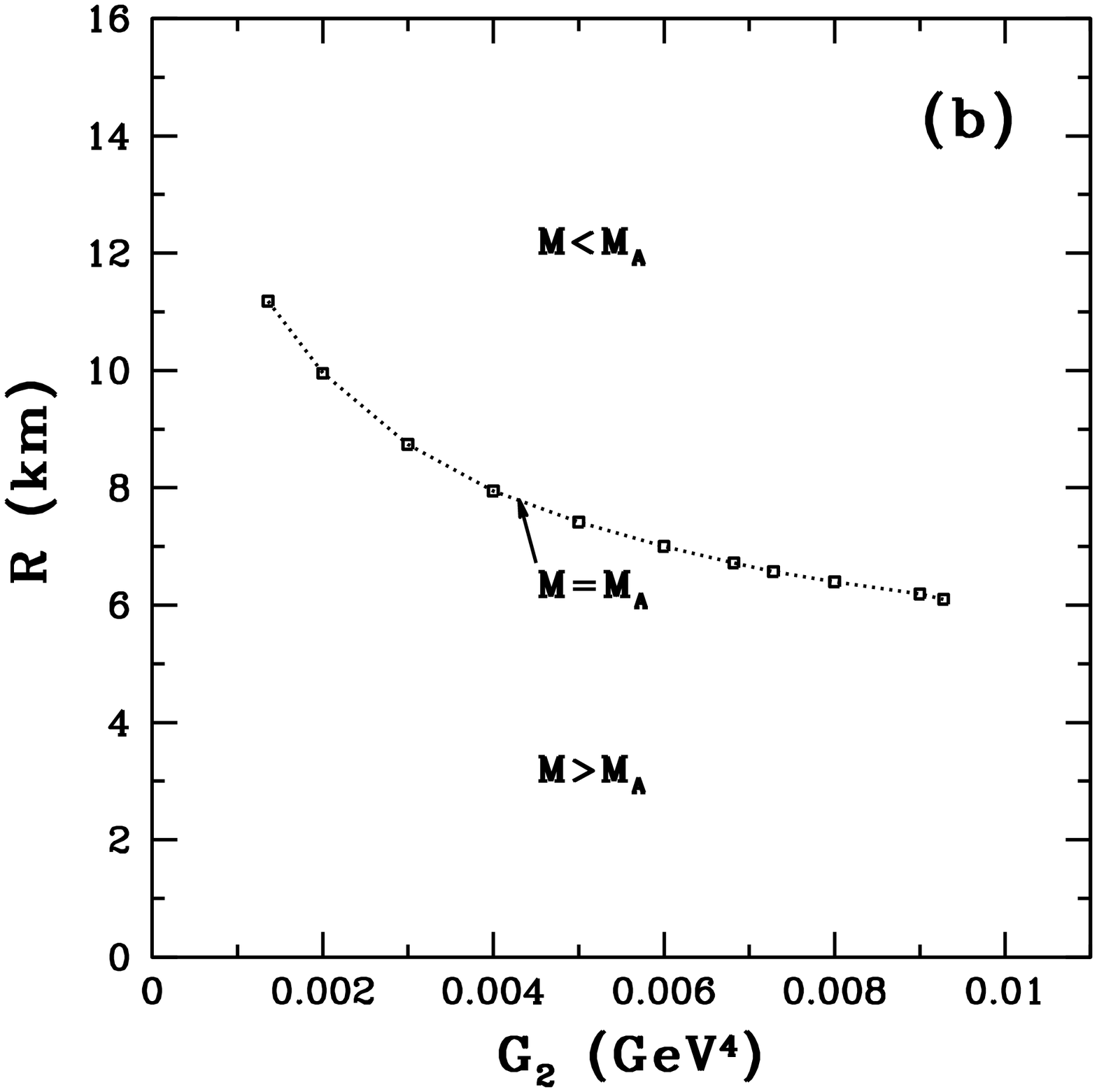,width=3.2truein,height=3.2truein}
\hskip .5in}
\caption{
Panel (a): Limiting masses of strange stars corresponding to the $M=M_A$ 
curve of Fig. \ref{g2v1} as function of $G_2$. 
Panel (b): The same as in Panel (a), but for limiting radii. 
In both panels, the stable sequences lie above the dotted curves.}
\label{g2m}
\end{figure*}

  Fig. \ref{g2m} shows the masses (in panel (a)) and radii (in panel (b)) for each point $(G_2,V_1)$ 
of Fig. \ref{g2v1}, as well as the $M<M_A$ and $M>M_A$ regions.  
 In both panels, along the $M=M_A$ curves we have masses in the range $1<M/M_\odot<2$ 
(panel (a)) and corresponding radii in the range 6 km$<R<$12 km (panel (b)).
 Additionally, for $G_2=0.00136\;{\rm GeV}^4$ and $V_1\simeq0.24\;{\rm GeV}$ we have $M/M_\odot=1.76$ 
and $R=11.2$ km. 
 In the other extreme, for $G_2=0.00927\;{\rm GeV}^4$ and $V_1=0$, we have $M/M_\odot=1.11$ and 
$R=6.1$ km.
 Finally, for $G_2=0.00682\;{\rm GeV}^4$ we have $M/M_\odot=1.2$ and $R=6.7$ km, showing that for 
the prescriptions of $G_2$ given by \cite{ST1,ST2} astrophysical masses and radii are allowed for 
$V_1$ in the range $0\leq V_1<0.0483$ GeV.
 
\subsubsection{Mass-radius relations}
\lb{mrr}

 Although it is very difficult to distinguish SS's from NS with respect to many 
observable phenomena, their MRR's present striking qualitative differences. 
 These differences are well illustrated in Fig. 3 of \cite{Bom2} which shows MRR's 
for several hadronic and SS star models. 
 NS generally have masses that increase from a minimum value with decreasing radii. 
 In contrast, there is no minimum mass for SS's and for $M<1M_\odot$, $M\propto R^3$.  
 On the other hand, for masses in the range $1M_\odot<M<2M_\odot$ the radii of SS's can be as large 
as 10 km. 
 However, this conclusions are valid for the MIT Bag model which has been largely used. This is not the 
case when we calculate MRR's in the framework of the FCM, as we show in what follows.

\begin{figure*}[th]
\centerline{
\psfig{figure=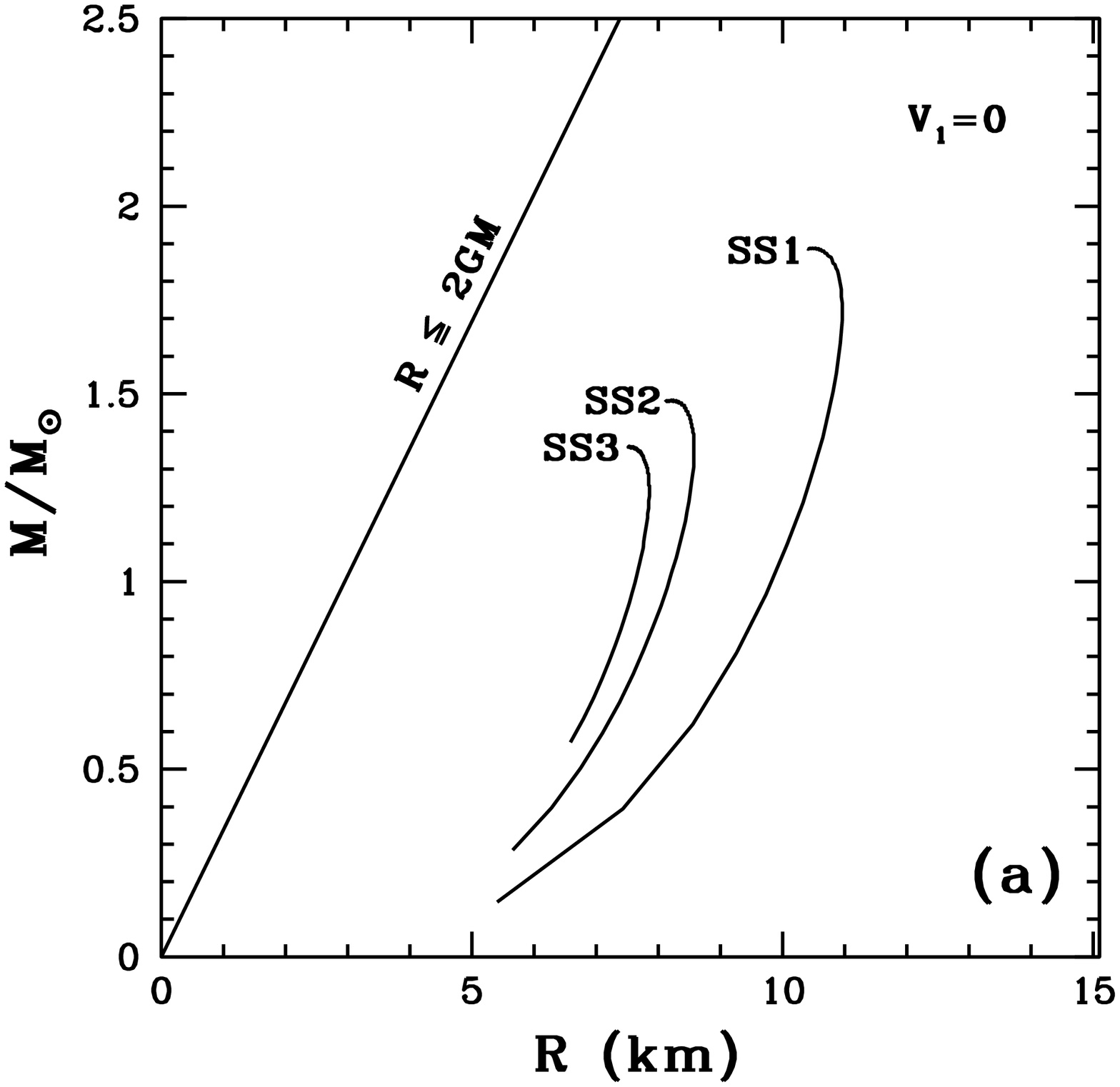,width=3.2truein,height=3.2truein}
\psfig{figure=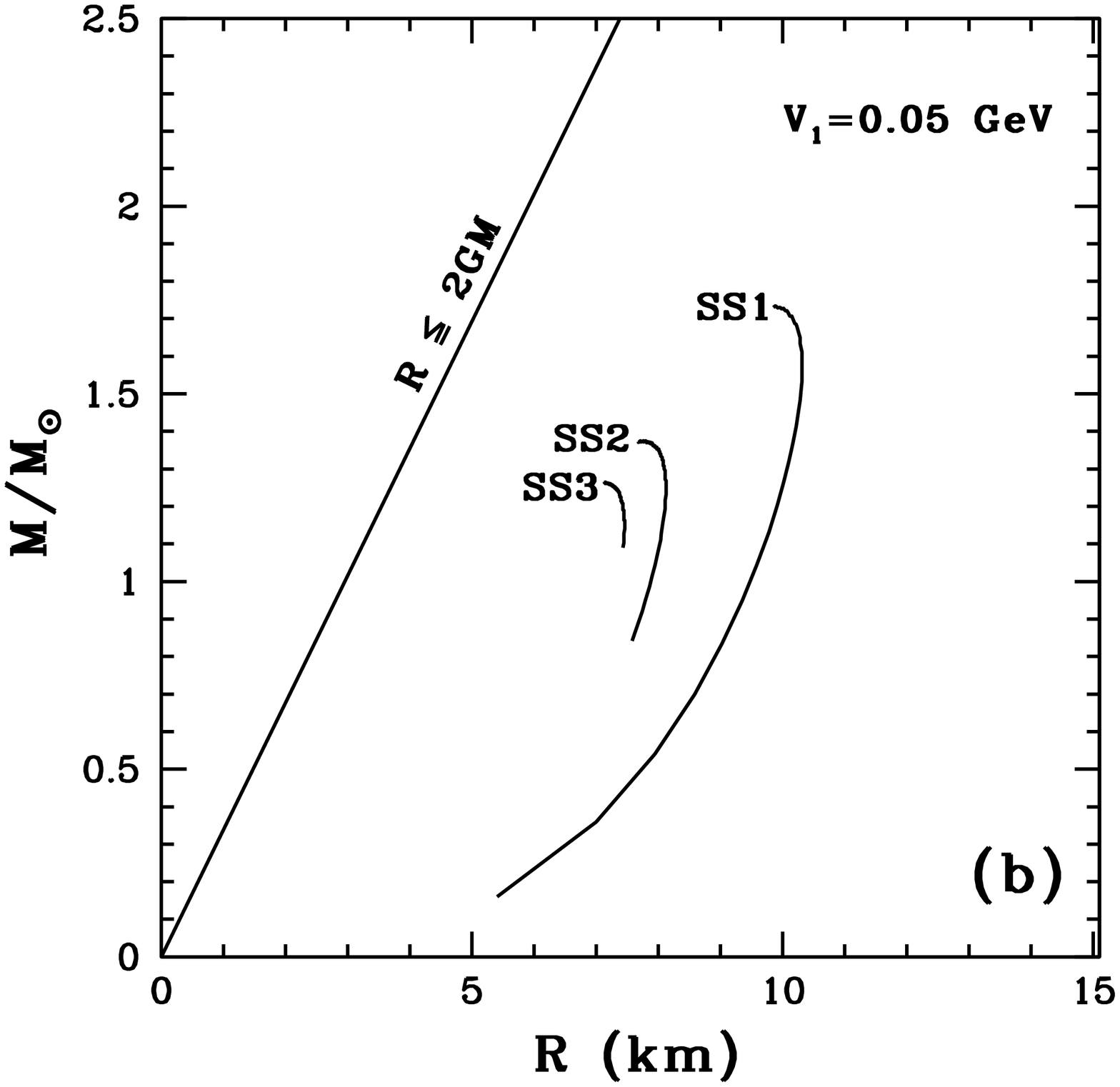,width=3.2truein,height=3.2truein}
\hskip .5in}
\caption{ 
The mass-radius relation for different values of $G_2$ 
and $V_1$. 
In both panels, the curves are labeled by SS1 for 
$G_1=0.003\;{\rm GeV}^4$, SS2 for $G_1=0.005\;{\rm GeV}^4$ and 
SS3 for $G_1=0.006\;{\rm GeV}^4$.
The solid straight line gives the Schwarzschild radius 
as function of stellar mass.}
\label{stmr1}
\end{figure*}

 To show the sensitivity of the theoretical predictions with respect to the FCM parameters we 
calculate straightforwardly the MRR's relations as done for $M-\varepsilon$ and $R-\varepsilon$ 
relations in previous sections. 
 In Fig. \ref{stmr1}, we display the variation of the MRR's for different values of $G_2$ and/or 
$V_1$. 
 The region for which $R\leq2GM$ is excluded to satisfy the condition that the radius $R$ must be 
larger than the black hole surface radius $2GM$. 
 The curves end near the maximum masses, where instability against gravitational collapse to black 
hole sets in. 
 In the left panel, depending on the values of $G_2$ (with $V_1=0$) the masses can be as 
large as $2M_\odot$ with radii around 11 km. 
 By switching on $V_1$, as shown in the right panel, we observe a similar trend to the case $V_1=0$, 
but with lower masses and radii.  
 Additionally, we observe that in contrast with the MIT Bag model the sequences for the FCM present lower 
limits which rise when $G_2$ and/or $V_1$ becomes larger. 

  Before we compare the FCM theoretical predictions with observations, let us summarize the 
characteristics of some observed sources taken as SS candidates.
  One of the best studied X-ray pulsars is Her X-1 discovered by Tananbaum te 
al. \cite{Tan} in 1972.  This source is generally classified as a Low Mass X-ray Binary 
(LMXB) with $M=0.98\pm0.12\;M_\odot$ and an estimated radius $R=6.7\pm1.2$ km \cite{LDW}. 

\begin{figure*}[th]
\centerline{
\psfig{figure=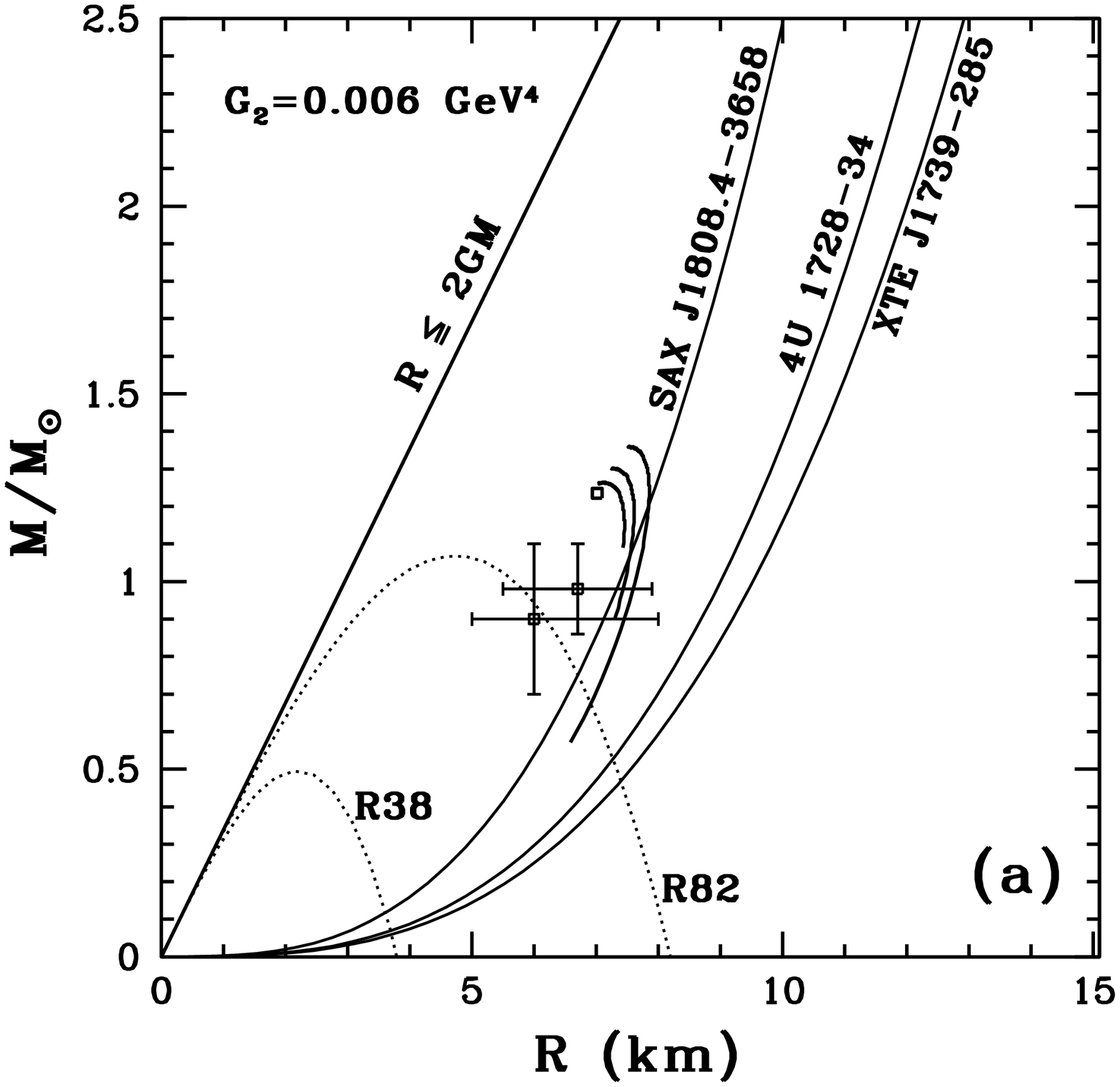,width=3.2truein,height=3.2truein}
\psfig{figure=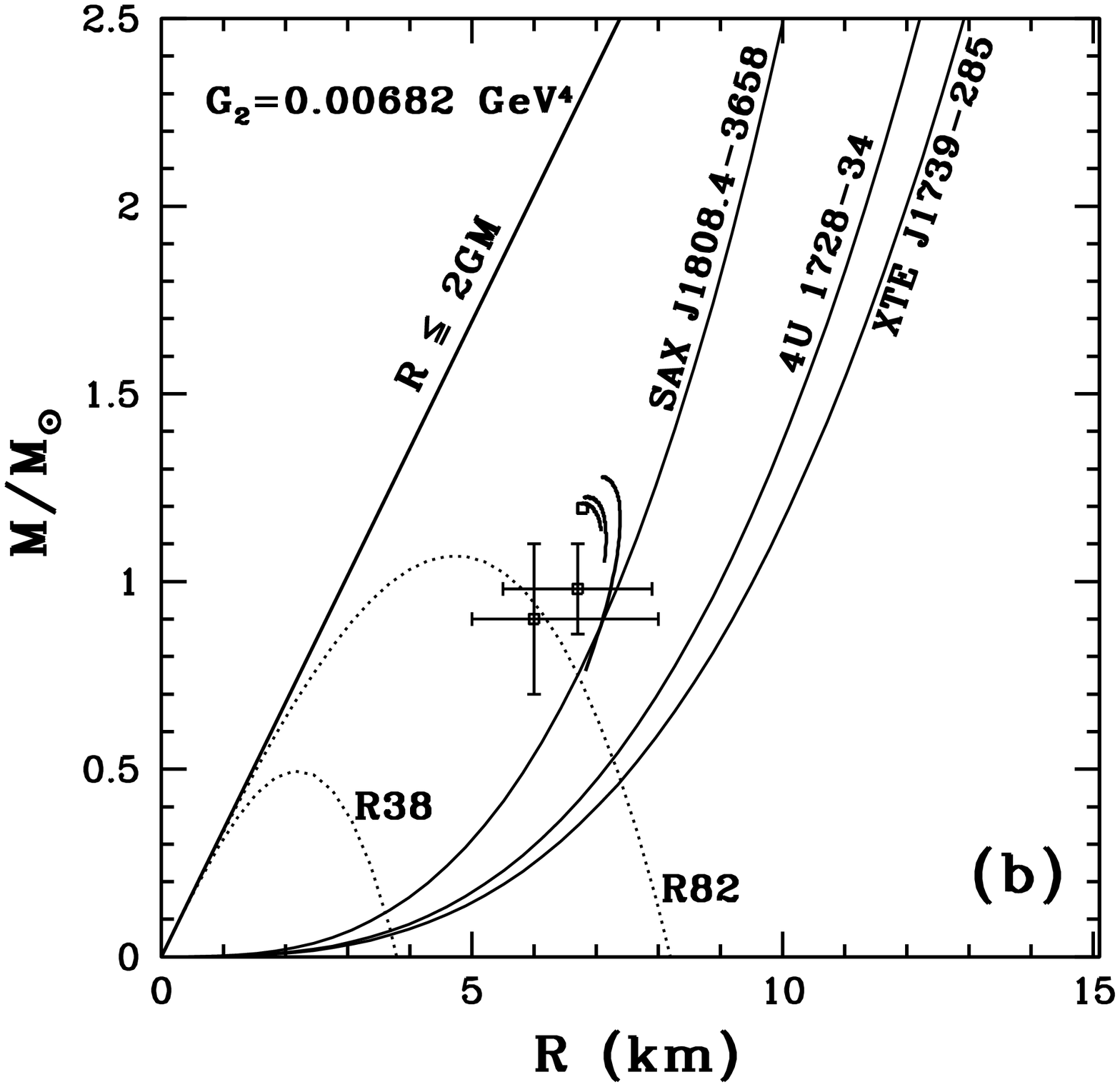,width=3.2truein,height=3.2truein}
\hskip .5in}
\caption{ 
The mass-radius relations for different values of model 
parameters compared with results extracted from observations. 
The crosses are the data of the sources Her X-1 and RX J1865.5-3754, respectively. 
The black solid curves give the upper limits for SAX J1808.4-3658, 
4U 1728-34 and XTE J1739-285.
The dotted curves indicate the radiation radii of RX J1865.5-3754, $R_\infty=3.8$ km (R38) 
$R_\infty=8.2$ km (R82).
The solid straight line gives the Schwarzschild radius as function of stellar mass. 
Panel (a): FCM results for $G_2=0.006{\rm GeV}^4$ and $V_1=0$ (red curve), 
$V_1=0.03$ GeV (blue curve), $V_1=0.05$ GeV (green curve), and $V_1=0.0671$ GeV (small open 
square dot) for the case of the limiting sequence with only a single star.
Panel (b): As in panel (a) but for $G_2=0.00682{\rm GeV}^4$ corresponding to 
$\Delta G_2=0.00341{\rm GeV}^4$ of Ref.\cite{Si6} and for $V_1=0$ (red curve), 
$V_1=0.03$ GeV (blue curve),  $V_1=0.04$ GeV (green curve), and $V_1=0.0483$ GeV (small open 
square dot ) as in panel (a).
For each of the sources SAX J1808.4-3658, 4U 1728-34 and RX J1865.5-3754, the  allowed regions 
for masses and radii are in between the Schwarzschild radius and the corresponding curve.}
\label{stmr3a}
\end{figure*}

 Another interesting strange star candidates is the nearby isolated X-ray source RX J1856.35-3754, 
from the deep Chandra LETG+HRC-S observation, with inferred mass $M=0.9\pm0.2M_\odot$ and radius 
$R=6^{-1}_{+2}$ km  \cite{Pons}. 
 This source has been object of controversial claims about its nature.  
 One group \cite{Dra} argue that it is a strange star whereas another group \cite{WaL} asserts that it 
is a normal neutron star.  
 The controversy arises mainly due to the distance estimate. 
 The main question concerns with the measurements of angular diameter, and thus the radiation radius. 
 Although it may be premature to conclude about the nature of this source, as pointed out in 
\cite{PrR}, we here follow the line of \cite{Dra} where RX J1856.35-3754 is best described as a SS. 
 One way to obtain information about the MRR is by the determination of the radiation radius of the 
source (see \cite{Dra} for details). 
 The inferred radiation radius $R_\infty$ is related to the mass $M$ and radius $R$ of the star by 
\bq
R_\infty=\frac{R}{\sqrt{1-2GM/R}}
\lb{Rinfty}
\eq
where the factor $R/\sqrt{1-2GM/R}$ comes from the gravitational redshift effect on the 
emitted radiation near the star surface. 
According to \cite{Dra} the inferred radiation radius is $R_\infty=3.8-8.2$ km for the 
distance to RX J1856.5-3754, $D=111-170$ pc.
  Rewriting  Eq. (\ref{Rinfty}) in terms of $M$ as function of $R$, we can obtain the relation 
between the mass and radius, for a fixed value of $R_\infty$. 
 In Fig. \ref{stmr3a} the MRR is represented by the dotted curves denoted by R38 and R82. 

  The source SAX J1808.4-3658, discovered with the Wide Field Camera on board the BeppoSAX 
satellite in September 1996 \cite{Zand}, is a transient X-ray source and LMXB.  
 From measured X-ray fluxes during the high- and low-states 
of the source, X.-D. Li et al. \cite{XDL1} have derived an upper limit for the stellar 
radius in SAX J1808.4-3658, which is given by
\bq
R\;\gappl\;28\big(\frac{F_{\rm max}}{F_{\rm min}}\big)^{-2/7}
\big(\frac{P}{2.49\;{\rm ms}}\big)^{-2/7}
\big(\frac{M}{M_\odot}\big)^{1/3}\;{\rm km}\;,
\lb{saxj1808}
\eq
where $F_{\rm max}$ and $F_{\rm min}$ are the X-ray fluxes measured during high- and low-states, 
respectively, and $P$ is pulse period. 
 The result is depicted in Fig.\ref{stmr3a}. 
 In \cite{XDL1} the results indicate that SS models are more consistent with 
the above semi-empirical MRR for SAX J1808.4-3658 than hadronic star models. 

  The transient X-ray XTE J1739-285 was observed by NASA Rossi X-ray Timing Explorer  
satellite \cite{Kaa}. It is the fastest-spinning celestial body yet known with a period 
$P=0.891$ ms. Using this rotation period, Lavagetto et al. \cite{Lav} derived an upper 
limit for the compact star radius in XTE J1739-285, given by
\bq
R\;<\; 9.52\big(\frac{M}{M_\odot}\big)^{1/3}\;{\rm km}\;,
\lb{xtej1739}
\eq
which is also shown in Fig.\ref{stmr3a}.

  The atoll source 4U 1728-34 is a LMXB observed with the Rossi X-Ray Timing Explorer (RXTE) 
which present kilohertz quasi-periodic oscillations (khz QPOs) \cite{HvK}.  
 Based on the works of Osherovich ant Titarchuk \cite{OsT,TOs}, Xiang-Dong Li et al. \cite{XDL2} 
have derived the following semi-empirical upper bound for 4U 1728-34 
\bq
R\;\gappl\;9\big(\frac{a_k}{1.03}\big)^{2/3}\big(\frac{M}{M_\odot}\big)^{1/3}\;{\rm km}\;,
\lb{4u1728}
\eq
where $a_k$ is given by Eq. (1) in \cite{XDL2}, and have shown that it is possibly a 
SS candidate.  
 In Fig. \ref{stmr3a}, the constraint given by Eq. (4) in \cite{XDL2} is not included in 
the MRR of 4U 1728-34.

 Finally, the binary millisecond pulsar J1614-2230, very recently observed with the 
Green Bank Telescope, show the measured mass $1.97\pm0.04\;M_\odot$ \cite{Dem}, which is 
the one with the best determination obtained to date. 
 However, it is described as a neutron star with an interior region, at a given baryon density, 
where the transition from nuclear matter to quark matter takes place \cite{OzD}, rather than 
properly a strange star candidate. 
 In the framework of our calculations, the mass of the J1614-2230 pulsar could only be achieved 
for $G_2\gappl 0.003\;{\rm GeV}^4$ as shown in Fig. \ref{stmr1} (curve labeled SS1), which is a very low value of 
$G_2$ with respect to that obtained from lattice calculations \cite{ST1,ST2}. 
 Given that it is hard to describe the J1614-2230 pulsar by hadronic or hybrid models, as in 
Fig. 3 of \cite{Dem}, this subject must be carefully investigated. 
However, this is not the scope of the present paper.

  We now compare the theoretical predictions for the MRR's of SS's, for several values of the FCM 
parameters, with the semi-empirical MRR's extracted from astrophysical observations. 
 The main resutls of our calculation are depicted in Fig. \ref{stmr3a} by red, blue and green 
curves, which show the MRR's for $G_2=0.006\;{\rm GeV}^4$ and $G_2=0.00682\;{\rm GeV}^4$ 
\cite{ST1,ST2} and for several values of the potential $V_1$.
 Generally speaking we can conclude that the FCM gives sequences of SS's within the allowed 
regions of the semi-empirical MRR's. 
 In both panels, the minimum masses span over a range between $\sim0.6\;M_\odot$ and 
$\sim1.12\;M_\odot$, and the maximum masses between $\sim1.2\;M_\odot$ and $\sim1.36\;M_\odot$. 
 The respective radii span over $\sim6.5$ km and $\sim7.8$ km. 
 For $G_2=0.006\;{\rm GeV}^4$ (panel (a)), the red and blue curves cross the 
error bars of Her X-1 and RX J1865.35-3754 indicating that sequences of SS's are possible 
for values of $V_1$ in the region $0\leq V_1<0.05$ GeV. 
 For $G_2=0.00682\;{\rm GeV}^4$ (panel (b)), the region of possibilities comes into the more 
restricted region $0\leq V_1<0.04$ GeV. 
 By increasing the values of $G_2$, the curves approximate to the central points of the error 
bars but the sequences become more and more restricted to lower values of $V_1$ due to the shrinking 
of the curves.  
 The extreme limit (although unrealistic with respect to the lattice calculations) corresponds to 
$G_2=0.009278\;{\rm GeV}^4$ and $V_1=0$ (see Fig. \ref{g2v1}) for which we obtain (see also Fig. 
\ref{g2m}) $M=1.11M_\odot$ and $R=6.1$ km, just in between the upper part of the error bars 
of the sources Her X-1 and RX J1865.35-3754. 
 This is the lower limit for maximum masses that can be calculated for all possible values of 
$G_2$ and/or $V_1$ in Fig. \ref{g2v1}.
 We observe that the results of the FCM are consistent with the theoretical MRR's for SS's shown in  
\cite{Koh,Bom1,Bom2,PrR,Zha}. 
 Then, from the above, we conclude that for values of $G_2$ obtained by comparison with lattice 
calculations on the critical temperature in \cite{ST1,ST2} it is possible to calculate strange 
star sequences within the regions allowed by 
the semi-empirical MRR's extracted from astrophysical observations and slightly consistent with the 
masses radii of Her X-1 and RX J1865.5-3754.

\subsection{The Bodmer-Witten conjecture}
\lb{witt}

  Since long ago, the question about the existence of an absolute ground state lower 
then that of normal nuclear matter has been considered .
 Bodmer and Witten pointed out a possibility that strange matter might be absolutely stable with 
respect to $^{56}F_{\rm e}$ at zero temperature and pressure \cite{Bod,Wit}. 
 A condition for the energy per nucleon is that ${\varepsilon}/n_B<m_N$ and a more stringent 
one is that strange quark matter be energetically preferred to the $^{56}F_{\rm e}$  
nucleus, which is the absolute ground state of cold matter at zero temperature and pressure, 
\bq
{\varepsilon}/n_B\;<\;M(^{56}F_{\rm e})/56\;,
\lb{fe56}
\eq
where $M(^{56}F_{\rm e})$ is the mass of the $^{56}F_{\rm e}$ nucleus.

\begin{figure*}[th]
\centerline{
\psfig{figure=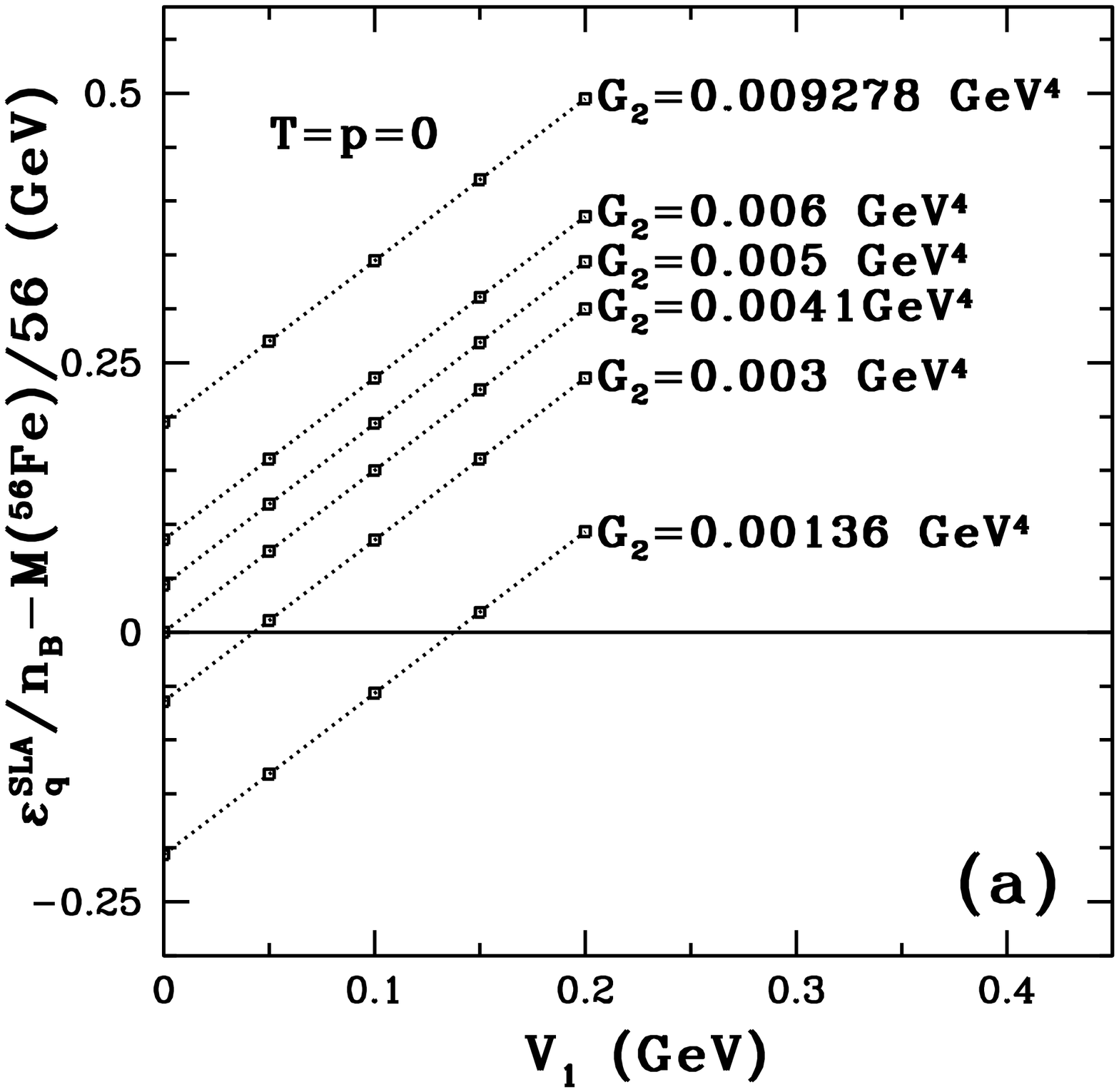,width=3.2truein,height=3.2truein}
\psfig{figure=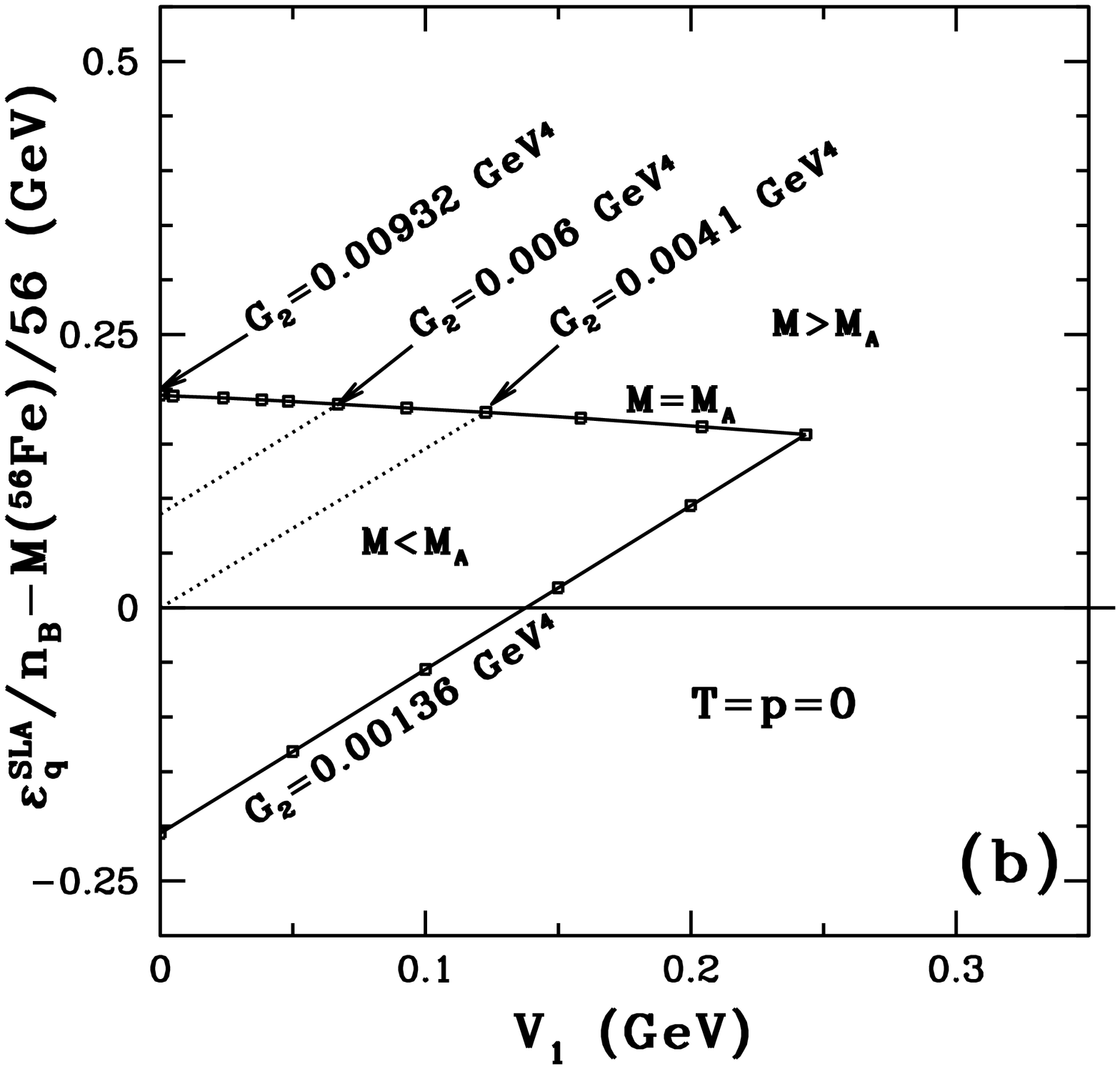,width=3.2truein,height=3.2truein}
\hskip .5in}
\caption{ 
Panel (a): Energy per baryon of electrically neutral strange matter in 
$\beta$-equilibrium minus the energy per nucleon $M(^{56}F_{\rm e})/56$ 
of $^{56}F_{\rm e}$ nucleus as function of $V_1$ calculated for different  
choices of $G_2$.
Panel (b): The same as in panel (a), but for the points of the $M=M_A$ curve in Fig. 
\ref{g2v1}, showing the regions $M<M_A$ and $M>M_A$. }
\label{ebt0p0}
\end{figure*}

  In the presente work, for some choices of the values of $G_2$ and $V_1$, the energy 
per baryon can be lower than the mass of the nucleon ($\sim939$ MeV) and  
the energy per nucleon in the $^{56}F_{\rm e}$ nucleus ($\sim930$ MeV).  
 In panel (a) of Fig. \ref{ebt0p0} we plot $\varepsilon^{SLA}_q/n_B-M(^{56}F_{\rm e})/56$ 
as function of $V_1$ for several values of $G_2$. 
 Due to the linearity of the energy per baryon as function of $V_1$, we have taken only a few 
points for the curves with constant $G_2$. 
 Depending on the values of $G_2$, two possibilities are allowed. 
 For $G_2\geq0.0041{\rm GeV}^4$ we have $\varepsilon^{SLA}_q/n_B\geq M(^{56}F_{\rm e})/56$ 
for all $V_1\geq0$. 
 When $G_2<0.0041{\rm GeV}^4$, we observe that $\varepsilon^{SLA}_q/n_B<M(^{56}F_{\rm e})/56$ 
for $0\leq V_1<V_1^{(0)}$ and  $\varepsilon^{SLA}_q/n_B\geq M(^{56}F_{\rm e})/56$ for 
$V_1\geq V_1^{(0)}$, where $V_1^{(0)}$ is the point at which 
$\varepsilon^{SLA}_q/n_B-M(^{56}F_{\rm e})/56$ crosses zero. 
 We explore this fact to investigate the possibility of the existence of absolutely 
stable quark matter with respect to $^{56}F_{\rm e}$ in strange stars.

 To this end, for each point ($G_2,V_1$) of the $M=M_A$ curve of Fig. \ref{g2v1}, we calculate 
the energy per baryon at $T=p=0$, which corresponds to quark matter conditions on star surfaces. 
 The results are shown in panel (b) of Fig. \ref{ebt0p0} where 
$\varepsilon^{SLA}_q/n_B-M(^{56}F_{\rm e})/56$ is plotted as function of $V_1$. 
 As in the panel (a), to each point of the $M=M_A$  curve corresponds a line of constant 
$G_2$ (not all shown in the right panel (b)).  
 All strange star sequences for which $M<M_A$ are inside the triangle. 
 We see that strange quark matter is unbound with respect to $^{56}F_{\rm e}$ for 
$G_2\geq0.0041\;{\rm GeV}^4$.
  On the other hand, we also see that $\varepsilon^{SLA}_q/n_B-M(^{56}F_{\rm e})/56<0$ 
is only possible for values of $G_2$ and $V_1$ in the regions 
$0.00136\;{\rm GeV}^4\leq G_2<0.0041\;{\rm GeV}^4$ and $0<V_1<0.138$GeV, respectively. 
 Thus, according to the FCM predictions for $G_2$, absolutely bound strange matter with 
respect to $^{56}F_{\rm e}$ on the surfaces of strange stars seems to be not allowed.

\section{Final remarks and conclusions}
\lb{frem}

 To summarize, we have investigated the main features of SQM with charge neutrality and 
in $\beta$-equilibrium and SS's in the framework of the FCM.
 In the FCM \cite{Si6}, the dynamics of confinement naturally appears in the EOS having 
the gluon condensate $G_2$ and the large distance ${\rm q\bar{q}}$ interaction potential 
$V_1$ as the main parameters of the model, for fixed values of the quark masses and temperature.
 
 We have considered the general aspects of the SQM at $T=0$. 
 The structure of the EOS, provided by the FCM, is such that it becomes softer (in the sense 
that for a given density we get lower pressure) with increasing $G_2$ and/or $V_1$.
 An interesting feature is that the energy per baryon is very sensitive to the values of $G_2$ 
and/or $V_1$ allowing for the investigation of the Bodmer-Witten conjecture \cite{Bod,Wit}.

 We calculate stable sequences of SS's for different values of $G_2$ 
and $V_1$. 
 The general feature is that the maximum masses of the sequences and the respective radii 
become lower, and the length of the sequences becomes shorter for larger values of $G_2$ 
and/or $V_1$.  
 Then, looking for appropriate values of $G_2$ and/or $V_1$, we obtain for limiting sequences with only one star 
lower bounds for the maximum masses, which determine a curve in the $G_2\times V_1$ plane below which all sequences are gravitationally stable. 
 Thus, independently of what the lattice prescriptions for the model parameters may be, 
we have obtained a set of values for $G_2$ and $V_1$ (which, of course, includes those obtained 
by comparison with lattice calculations \cite{ST1,ST2}) which produces stable SS sequences. 
 The parameters are constrained by astrophysical conditions to spun over the ranges 
$0<G_2<0.00927\;{\rm GeV}^4$ and $0\leq V_1<V_1^{max}$, where $V_1^{max}$ is an upper bound 
determined by the above curve in the $G_2\times V_1$ plane. 
 This gives us a general connection that open the possibilities for comparison of lattice data with 
results obtained from astrophysical observations.
 
 In going farther, we analyze the sensitivity of the MRR's with respect to the variation of 
the model parameters. 
 The MRR's become shorter and with lower masses and radii with increasing values of 
$G_2$ and/or $V_1$ and, depending on the values of these parameters, it could reach upper 
masses around $\sim2{\rm M}_\odot$ and the respective radii around $\sim10$ km.
 In order to make a more specific connection between nonperturbative QCD and astrophysics, we 
compare the theoretical predictions of masses and radii of SS's provided by the FCM with those 
extracted from astrophysical observations of some SS candidates.  
 For $G_2=0.006\;{\rm GeV}^4$ and $G_2=0.00682\;{\rm GeV}^4$ (extracted from lattice predictions 
for the critical temperature at zero chemical potential \cite{ST1,ST2}), we obtain stable sequences 
within the allowed regions of the semi-empirical MRR's of the sources  
SAX J1808.4-3658, 4U 1728-34 and XTE J1739-285, and slightly compatible with the mass and radius of 
the sources Her X-1, RX J1865.5-3754.
 We remark that it seems possible to reconcile astrophysical results with the predictions 
for the gluon condensate in \cite{ST1,ST2}.

 Finally, we have considered the Bodmer-Witten \cite{Bod,Wit} conjecture in the framework of the FCM. 
 We have calculated the energy per baryon of SQM with charge neutrality and $\beta$-equilibrium 
at zero temperature and pressure for different values of $G_2$ and $V_1$.
 We have found that only for $G_2<0.0041\;{\rm GeV}^4$ the energy per baryon is less than the bind 
energy per nucleon of the $^{56}F_{\rm e}$ nucleus. 
 Thus, for $G_2=0.006-0.007\;{\rm GeV}^4$ absolutely bound SQM with respect to $^{56}F_{\rm e}$ on 
the SS surfaces seems to be not possible, unless future developments show to be otherwise.

\newpage

%
%

\newpage


\newpage


\end{document}